\documentclass[twocolumn,twocolappendix]{aastex631}

\usepackage{multirow}
\usepackage{graphicx}

\shorttitle{Sulfur and Multiple Molecules on GJ 3470 b}
\shortauthors{Beatty et al.}

\begin{document}

\newcommand{\bjdtdb}{${\rm {BJD_{TDB}}}$}
\newcommand{\feh}{{\left[{\rm Fe}/{\rm H}\right]}}
\newcommand{\teff}{{T_{\rm eff}}}
\newcommand{\ecosw}{${e\cos{\omega_*}}$}
\newcommand{\esinw}{${e\sin{\omega_*}}$}
\newcommand{\msun}{${\rm M}_\odot$}
\newcommand{\rsun}{${\rm R}_\odot$}
\newcommand{\lsun}{${\,{\rm L}_\Sun}$}
\newcommand{\mj}{${\,{\rm M}_{\rm J}}$}
\newcommand{\rj}{${\,{\rm R}_{\rm J}}$}
\newcommand{\me}{${\,{\rm M}_{\oplus}}$}
\newcommand{\re}{${\,{\rm R}_{\oplus}}$}
\newcommand{\fave}{\langle F \rangle}
\newcommand{\fluxcgs}{10$^9$ erg s$^{-1}$ cm$^{-2}$}
\newcommand{\three}{3.6\,$\mu$m\ }
\newcommand{\four}{4.5\,$\mu$m\ }
\newcommand{\threealt}{3.6\,$\mu$m}
\newcommand{\fouralt}{4.5\,$\mu$m}
\newcommand{\um}{$\mu$m}
\newcommand{\water}{$\mathrm{H}_2\mathrm{O}$}
\newcommand{\methane}{$\mathrm{CH}_4$}
\newcommand{\carbondioxide}{$\mathrm{CO}_2$}
\newcommand{\sulfurdioxide}{$\mathrm{SO}_2$}
\newcommand{\hydrogensulfide}{$\mathrm{H}_2\mathrm{S}$}
\newcommand{\JWST}{\project{JWST}}                               
\newcommand{\HST}{\project{HST}} 
\newcommand{\Spitzer}{\project{Spitzer}}                            
\newcommand{\Kepler}{\project{Kepler}}  
\newcommand{\TESS}{\project{TESS}}  

\newcommand{\red}{\textcolor{red}}

\title{Sulfur Dioxide and Other Molecular Species in the Atmosphere of the Sub-Neptune GJ 3470 b}

\author[0000-0002-9539-4203]{Thomas G. Beatty}
\affiliation{Department of Astronomy, University of Wisconsin--Madison, Madison, WI, USA}

\author[0000-0003-0156-4564]{Luis Welbanks}
\affiliation{School of Earth and Space Exploration, Arizona State University, Tempe, AZ, USA}

\author[0000-0001-8291-6490]{Everett Schlawin}
\affiliation{Steward Observatory, University of Arizona, Tucson, AZ, USA}

\author[0000-0003-4177-2149]{Taylor J. Bell}
\affiliation{Bay Area Environmental Research Institute, NASA's Ames Research Center, Moffett Field, CA, USA}

\author[0000-0001-6247-8323]{Michael R. Line}
\affiliation{School of Earth and Space Exploration, Arizona State University, Tempe, AZ, USA}

\author[0000-0002-8517-8857]{Matthew Murphy}
\affiliation{Steward Observatory, University of Arizona, Tucson, AZ, USA}

\author[0000-0001-8745-2613]{Isaac Edelman}
\affiliation{Bay Area Environmental Research Institute, NASA's Ames Research Center, Moffett Field, CA, USA}
\affiliation{Space Science and Astrobiology Division, NASA's Ames Research Center, Moffett Field, CA, USA}

\author[0000-0002-8963-8056]{Thomas P. Greene}
\affiliation{Space Science and Astrobiology Division, NASA’s Ames Research Center, Moffett Field, CA, USA}

\author[0000-0002-9843-4354]{Jonathan J. Fortney}
\affiliation{Department of Astronomy and Astrophysics, University of California, Santa Cruz,
1156 High Street, Santa Cruz, CA 95064, USA}

\author[0000-0003-4155-8513]{Gregory W. Henry}
\affiliation{Tennessee State University, Nashville, TN, USA (retired)}

\author[0000-0003-1622-1302]{Sagnick Mukherjee}
\affiliation{Department of Astronomy and Astrophysics, University of California Santa Cruz, Santa Cruz, CA, USA}

\author[0000-0003-3290-6758]{Kazumasa Ohno}
\affiliation{Division of Science, National Astronomical Observatory of Japan, Tokyo, Japan}
\affiliation{Department of Astronomy and Astrophysics, University of California Santa Cruz, Santa Cruz, CA, USA}

\author[0000-0001-9521-6258]{Vivien Parmentier}
\affiliation{Laboratoire Lagrange, Observatoire de la Côte d’Azur, Université Côte d’Azur, Nice, France}

\author[0000-0003-3963-9672]{Emily Rauscher}
\affiliation{Department of Astronomy, University of Michigan, Ann Arbor, MI, USA}

\author[0000-0002-3295-1279]{Lindsey S. Wiser}
\affiliation{School of Earth and Space Exploration, Arizona State University, Tempe, AZ, USA}

\author[0000-0002-3034-8505]{Kenneth E. Arnold}
\affiliation{Department of Astronomy, University of Wisconsin--Madison, Madison, WI, USA}

\begin{abstract}

We report observations of the atmospheric transmission spectrum of the sub-Neptune exoplanet GJ 3470 b taken using the Near-Infrared Camera (NIRCam) on JWST. Combined with two archival HST/WFC3 transit observations and fifteen archival Spitzer transit observations, we detect water, methane, sulfur dioxide, and carbon dioxide in the atmosphere of GJ 3470 b, each with a significance of $>3\,\sigma$. GJ 3470 b is the lowest mass -- and coldest -- exoplanet known to show a substantial sulfur dioxide feature in its spectrum, at $M_{p}$=11.2\,\me\ and $T_{eq}$=600\,K. This indicates disequilibrium photochemistry drives sulfur dioxide production in exoplanet atmospheres over a wider range of masses and temperatures than has been reported \citep{tsai2023so2} or expected \citep{polman2023so2}. The water, carbon dioxide, and sulfur dioxide abundances we measure indicate an atmospheric metallicity of approximately $100\times$ Solar. We see further evidence for disequilibrium chemistry in our inferred methane abundance, which is significantly lower than expected from equilibrium models consistent with our measured water and carbon dioxide abundances.  

\end{abstract}

\section{Introduction}

One of the surprising results from exoplanet demographics studies is the prevalence of Neptune- and sub-Neptune-sized exoplanets \citep{fulton2018radgap}. Given their ubiquity, understanding how these exoplanets form is an important part of understanding planet formation. One way to do this is by studying their atmospheric compositions, since the carbon, oxygen, and overall metal abundances in Neptune- and sub-Neptune-mass planetary atmospheres are expected to be strongly influenced by their formation histories \citep{mordasini2016mhpredictions}. Existing observations of these atmospheres from HST provide some water abundance measurements \citep{wakeforf2017hatp26,morley2017gj436,benneke2019k218}, but the difficulty in making these measurements and the lack of detections of carbon-bearing molecules has made comparing to planet formation theories difficult \citep{welbanks2019mhtrends}. Recent JWST observations of K2-18b \citep{Madhu2023K218} and TOI-270d \citep{Benneke2024TOI270,Holmberg2024TOI270} have begun to reveal spectra with strong molecular features in this regime, though many other sub-Neptunes have appeared featureless \citep{LustigYaeger2023,May2023,Kirk2024}.

One of the canonical sub-Neptunes in this region of parameter space that is known to have features in its transmission spectrum is GJ 3470 b, which has a radius of 4.2\,\re, a mass of 11.2\,\me, and an equilibrium temperature of 600\,K \citep{bonfils2012}. This makes GJ 3470 b a relatively cool and relatively low-density member of the sub-Neptune population. Ground-based transit observations identified a Rayleigh-scattering slope in the atmosphere \citep{crossfield2013gj3470,dragomir2015gj3470} and indicated that the atmosphere was likely methane-depleted \citep{crossfield2013gj3470}. HST and Spitzer observations of the transmission spectrum of GJ 3470 b \citep{benneke2019gj3470} have shown significant water absorption near 1.4\um. These observations again indicated that the planetary atmosphere is likely depleted in methane, and the analysis also suggested GJ 3470 b has an atmospheric metallicity close to that of the Sun (as measured via the water abundance). These HST and Spitzer observations also suggested a possibly strong slope in the transmission spectrum between 2.5\um\ and 3.5\um, which was hypothesized to be caused by Mie scattering in the atmosphere.

The apparent low metallicity of GJ 3470 b has been a puzzle, since both Uranus and Neptune have well-measured metallicities at approximately 100 times the Solar value \citep{atreya2016ssmhtrends}. Planet formation theories also suggest that the metallicity of a lower-mass planet like GJ 3470 b should be 200 to 300 times Solar \citep{mordasini2016mhpredictions}. However, it is important to note that estimates for the metallicities of Solar System planets are based on methane-to-hydrogen ratios, while most previous metallicity estimates for exoplanets (including GJ 3470 b) have primarily used water-to-hydrogen ratios \citep{welbanks2019mhtrends}. This difference in metallicity estimation methods has made detailed comparisons between Solar System and exoplanet results difficult and has been necessitated by the lack of spectroscopic coverage of strong, isolated, methane absorption features in existing exoplanet atmosphere observations from either the ground or space.

GJ 3470 b is also an interesting target for atmospheric characterization observations because it appears to be significantly less cloudy than similar-mass exoplanets orbiting bright stars. Specifically, the other two ``canonical'' exoplanets in this mass range, GJ 436 b and GJ 1214 b, both show infrared transmission spectra that are consistent with flat lines in both HST/WFC3 \citep{knutson2014gj436,kreidberg2014gj1214} and more recent JWST/MIRI \citep[in the case of GJ 1214 b,][]{gao2023gj1214} observations, indicating that both have high-altitude aerosols around their terminators. GJ 3470 b, in contrast, shows significant features in its HST/WFC3 transmission spectrum \citep{benneke2019gj3470}, indicating that its clouds are lower in the atmosphere, more patchy, or some combination of the two. The reason for the differing levels of cloudiness between these three -- otherwise similar -- exoplanets is not clear at this time. 

The short period of the planetary orbit (3.337 days) places GJ 3470 b at the outer edge of the Neptune Desert, indicating that it may have had a dynamic migration and evolutionary history at earlier times. Transit observations of the Ly-$\alpha$ \citep{bourrier2018LyAlpha} and He I \citep{palle2020HeEscape} lines show that the planet is currently losing $\sim10^{10}$\,g\,s$^{-1}$ into space. This mass-loss rate would imply that GJ 3470 has lost roughly 40\% of its mass over the 2 Gyr system lifetime \citep{bourrier2018LyAlpha}. Rossiter-McLaughlin measurements of the spin-orbit alignment of the system \citep{Gumi2022PolarOrbit} show that GJ 3470 b is on a nearly-polar orbit ($\lambda = 98^{+15}_{-12}$\,deg.). Spitzer secondary eclipse observations \citep{benneke2019gj3470} indicate the orbit is not circular, and coupled with radial velocity observations the most recent analyses give $e=0.125\pm0.04$ \citep{kosiarek2019,Gumi2022PolarOrbit}.

The combination of the present-day high mass-loss rate, polar orbit, and eccentric orbit point towards the scenario that a more massive version of GJ 3470 b was dynamically scattered inwards towards the star GJ 3470 at some point after the planet formed \citep{Gumi2022PolarOrbit}, and subsequently lost a significant fraction of its outer envelope \citep{palle2020HeEscape}. The planet as we see it today is likely in the final stages of tidal circularization and atmospheric mass-loss on its way ``out'' of the Neptune Desert at the end of this turbulent migration history. 

Measurements of GJ 3470 b's present-day atmospheric carbon, oxygen, and sulfur abundances are one method to constrain the planet's initial formation location \cite{turrini2021so2formation,crossfield2023so2formation}, and to better understand the formation and migration histories of sub-Neptunes. 

\section{Observations, Data Analysis, and Fitting}

In our analysis of the transmission spectrum of GJ 3470 b, we used a combination of new data collected by JWST/NIRCam and archival transit observations from HST/WFC3 and Spitzer/IRAC. The WFC3 transit observations and some Spitzer data have been previously analyzed and reported on in \cite{benneke2019gj3470} For this work, we re-extracted and re-fit all of the archival data to minimize possible offsets in the overall transit depths between the different observations. These offsets are often caused by using different priors on the planetary orbital parameters and stellar limb-darkening coefficients, and we wished to fit all of the available data using a single set of orbital parameters and a set of limb-darkening coefficients drawn from a single stellar model.

\begin{figure*}[t!] 
\begin{center}                                 
\includegraphics[width=1.0\textwidth]{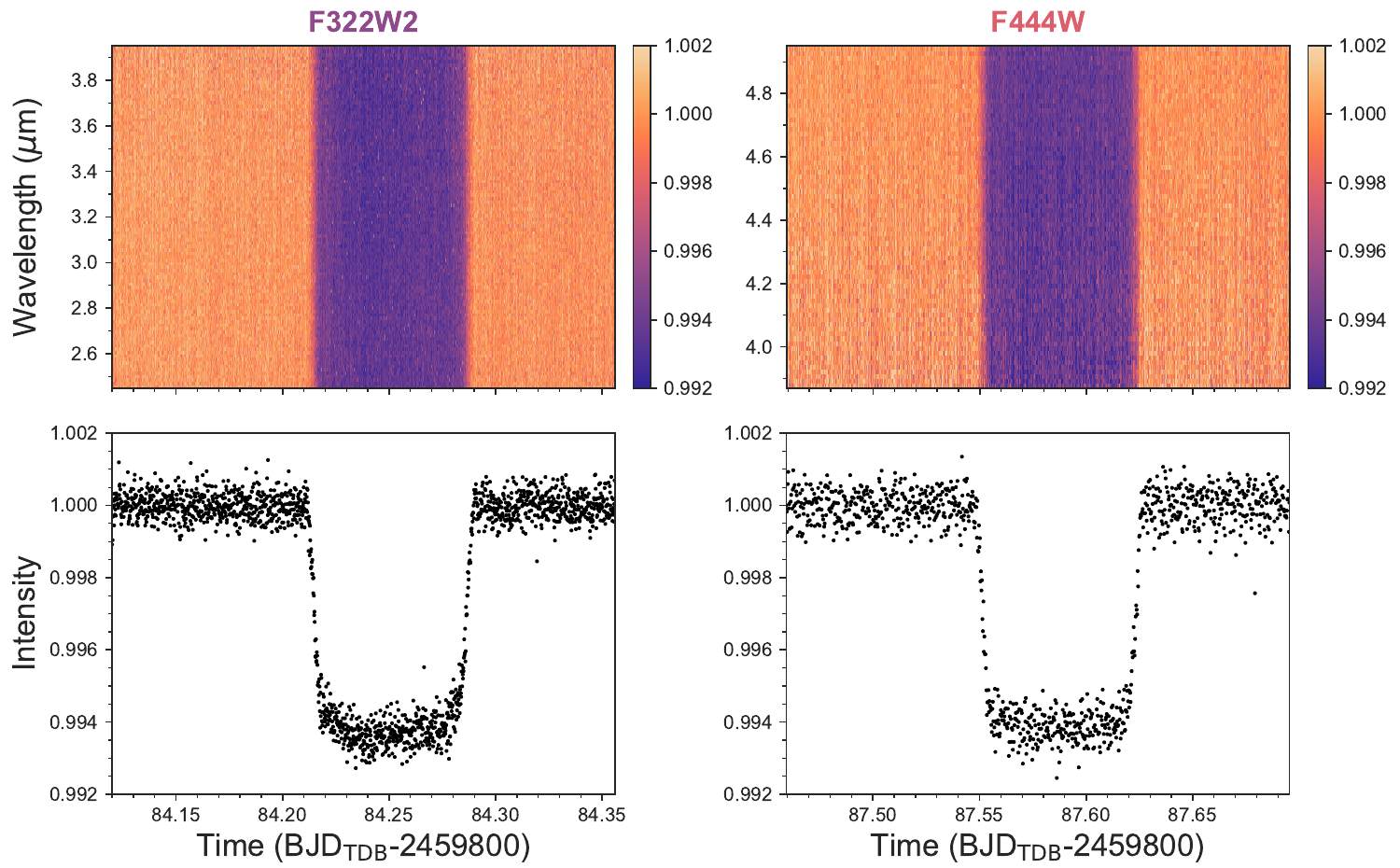}
\end{center}
\vskip -0.2in 
\caption{The spectroscopic and broadband NIRCam lightcurves of GJ 3470 b. The top panels show the raw F322W2 and F444W lightcurves binned in the spectral direction into 0.015\um-wide spectral bins, with the transit of GJ 3470 b clearly visible at all wavelengths. The bottom panels show the broadband F322W2 and F444W lightcurve data, without their associated error bars for visual clarity.}
\label{fig:waterfall}  
\end{figure*}

In addition to the transit observations, we used the Gaia parallax \citep{GaiaDR3} for GJ 3470 to fit the stellar spectral energy distribution (SED) and, hence, the stellar radius. Together with the stellar density we measure via the transit lightcurves and previous radial velocity observations \citep{kosiarek2019}, this allowed us to make new estimates for the absolute radii and masses for the star GJ 3470 and the planet GJ 3470 b. We used the values we determined for GJ 3470 b's radius and mass in our atmospheric retrievals.

We also update the results from long-term photometric monitoring of the variability of GJ 3470. Previous results estimated the star GJ 3470's rotation period and variability amplitude \citep{biddle2014} based on one year of monitoring data. We now have eleven years of photometric monitoring data from the Tennessee State University automated imaging telescope, which allowed us to refine the rotation period measurement and the variability amplitude at this period. As described below, based on previous analyses \citep{benneke2019gj3470}, the stellar variability we observe does not significantly influence our results.

We briefly summarize all of these steps here. Complete details and a fuller description of the data reduction, lightcurve fitting, and stellar SED and variability analyses are given in Appendices A and B.

\subsection{Observations and Data Reduction}

\subsubsection{JWST NIRCam}

We observed two transits of GJ 3470 b using JWST/NIRCam as a part of GTO program 1185 \citep[PI Thomas Greene,][]{manatee1}. We observed the first transit on UT 31 October 2022, using the F322W2 filter together with the \textsc{grismr} disperser on NIRCam's longwave (LW) detector, and the F210M filter together with the \textsc{wlp8} weak lens array on NIRCam's shortwave (SW) detector. For the F322W2 transit, we configured both detectors to use the \textsc{subgrism256} subarray in \textsc{bright2} readout mode with four readout amplifiers and using four groups per integration. Both the LW and SW detectors took 1681 integrations in the complete exposure. In total, we observed the first transit of GJ 3470 b for 5.67 hours. 

\begin{figure*}[t!] 
\begin{center}                                 
\includegraphics[width=1.0\textwidth]{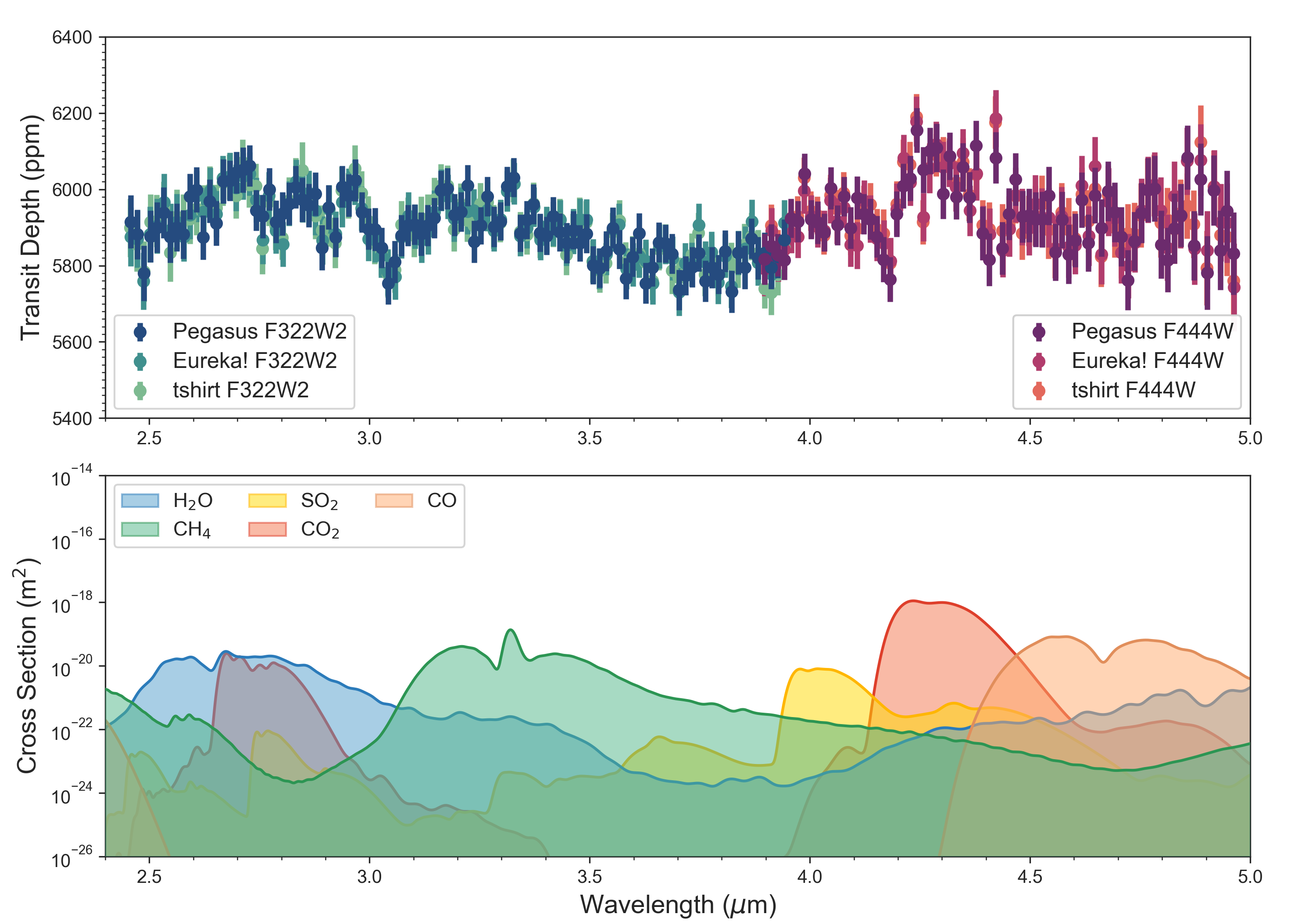}
\end{center}
\vskip -0.2in 
\caption{Three independent reductions of the NIRCam F322W2 and F444W data agree within their uncertainties The top panel shows the F322W2 and F444W transmission spectra, as reduced using the \texttt{Pegasus}, \texttt{Eureka}, and \texttt{tshirt} pipelines, plotted with their associated $1\,\sigma$ uncertainties using constant 0.015\um-wide spectral bins. All of the F322W2 spectra have been offset down by 163\,ppm to correct for a bias offset in the F322W2 transit depths compared to the F444W and Spitzer data (see Appendix B.3). All three reductions agree on the shape of the transmission spectrum and show features due to \water\ (2.5--3.2\um), \methane\ (3.2--3.5\um), \sulfurdioxide\ (3.9--4.1\um), and \carbondioxide\ (2.6--2.9\um\ and 4.2--4.5\um). The absorption cross sections of the prominent species in this wavelength range, many of which are visually identifiable in the bottom panel.}
\label{fig:spectrum}  
\end{figure*}

We observed the second transit of GJ 3470 b on UT 3-4 November 2022 using the F444W filter on the LW detector. The LW and SW detectors were configured similarly to the F322W2 transit, except that for this second transit, we increased the number of groups per integration to seven and decreased the number of integrations per exposure to 1009. This longer integration time kept the peak counts from GJ 3470 in the F444W filter roughly the same as in the F322W2 data. Again, we observed this second transit for 5.67 hours.

Our fiducial NIRCam data reduction used the \texttt{Pegasus} pipeline (\url{https://github.com/TGBeatty/PegasusProject}), which was initially developed for the NIRCam commissioning observations. To ensure the reproducibility of our results, we performed two additional analyses of the NIRCam spectra using the \texttt{Eureka!} \citep{bell2022eureka}, and \texttt{tshirt} \citep{ahrer2023wasp39,bell2023_methane} data reduction pipelines. Using the same system parameters, limb-darkening coefficients, and a freely-fit linear background trend, we find strong agreement between all three spectra, with differences of $\chi^2/\mathrm{dof}=0.13$ for \texttt{Pegasus} vs. \texttt{Eureka!} and $\chi^2/\mathrm{dof}=0.21$ for \texttt{Pegasus} vs. \texttt{tshirt}. The three spectra all show the same spectroscopic features with no major systematic differences between them (Figure \ref{fig:spectrum}). We chose the \texttt{Pegasus} spectrum as our fiducial case for atmospheric modeling due to its ability to reduce the archival HST and Spitzer data. The complete details of all three reduction pipelines and their results are described in Appendices A and B. 

\subsection{HST WFC3}

HST observed three transits of GJ 3470 b using Wide-Field Camera 3 (WFC3) and the G141 disperser as a part of GO program 13665 (PI Benneke) using four orbits of HST time per transit \citep{benneke2019gj3470}. To provide consistency with our analysis and assumed system and orbital parameters, we re-reduced and re-fit two of the transit observations: those taken on UT 13 March 2015 and UT 22 October 2015. We reduced the spectroscopic observations and extracted broadband and spectroscopic lightcurves using previously described techniques \cite{beatty2017kepler13}. The broadband lightcurve covered the wavelength range from 1.12\um\ to 1.66\um\ in a single channel, while the spectroscopic lightcurves covered this same wavelength range in fifteen bins that were each 0.036\um\ wide. This wavelength range and the spacing match the analysis of these data in \citep{benneke2019gj3470}.

\subsection{Spitzer IRAC}

The GJ 3470 system has been repeatedly observed in transit and eclipse by Spitzer at \three\ and \fouralt. The combined results from three of the \three transits and three of the \four transits have also been described in \cite{benneke2019gj3470}. Here, we reduced and simultaneously fit a total of seven \three and eight \four Spitzer transit observations of GJ 3470 (which includes the three in each channel that are already in the literature). These data were taken as a part of Spitzer GO programs 90092 (PI Desert), 80261 (PI Demory), and 13140 (PI Kreidberg). All of these observations used the same instrument and exposure time settings in subarray mode with 0.4-second exposures and PCRS peak-up. We reduced these data and extracted photometry using previously described procedures \citep{beatty2018,beatty2019}.

\subsection{Broadband and Spectroscopic Lightcurve Fitting}

We first performed a simultaneous fit to broadband HST, broadband JWST, and Spitzer transit lightcurves to refine the general orbital and planetary parameters of the GJ 3470 system. Our results from this joint fit to the broadband photometry (Table 2, Figure \ref{fig:broadband}) were generally consistent with previous measurements \citep{biddle2014,awiphan2016,benneke2019gj3470,kosiarek2019}. We did find that the planetary orbit was slightly less eccentric and the transit depth was mildly shallower \citep{benneke2019gj3470,kosiarek2019}, due to our analysis with the new NIRCam data measuring GJ 3470 b's impact parameter to be $b=0.31\pm0.02$, which is slightly smaller than those older estimates of $b\approx0.4$ \citep{dragomir2015gj3470,awiphan2016,benneke2019gj3470,kosiarek2019}.

We measured the spectroscopic transit depths of GJ 3470 b using the JWST/NIRCam observations and the archival HST/WFC3 data. For these fits, we fixed all of the transit properties except for the planet-to-star radius ratio to the values determined from the joint photometric fit. The JWST data was relatively clean and preferred a simple linear background temporal trend to remove systematic trends in the lightcurves. The HST/WFC3 data showed the usual hook-like systematic trends. We fit these by trending the data within each orbit using a hooked exponential and a visit-long linear slope, the details of which are described in more detail in \cite{beatty2017kepler13}. We found the mean depth of the WFC3 spectrum to be mildly shallower than previously reported in \cite{benneke2019gj3470}, a result of more refined measurements of $a/R_*$ and $\cos(i)$ driven by the JWST data. As described in Appendix B.3, we found that the F322W2 spectrum was offset $163\pm34$\,ppm high relative to the F444W and Spitzer transit depths. \cite{Welbanks2024w107} recently found a roughly similar offset between F322W2 and F444W transmission spectra of WASP-107 b, and \cite{xue2024hd209458} marginally detected an offset in their F322W2 and F444W transmission spectra of HD 209458 b. Possibly, the offset is due to flux-dependent non-linearity effects in the F322W2 observations in all three data sets -- though more investigation is necessary to confirm the cause of the offsets between F322W2 and F444W transit data.

\subsection{Stellar SED Modeling}

We used a set of catalog 2MASS JHK \citep{2mass} and AllWISE W1 and W2 \citep{wise} magnitudes for the GJ 3470 system to fit a spectral energy distribution (SED) model to the stellar emission. In conjunction with the Gaia \citep{GaiaDR3} parallax for the system, this allowed us to estimate a stellar radius for the star GJ 3470. To model the SED, we used BHAC15 spectra \citep{bhac15} for the stellar SED.

\begin{figure*}[t!] 
\begin{center}                                 
\includegraphics[width=1.0\textwidth]{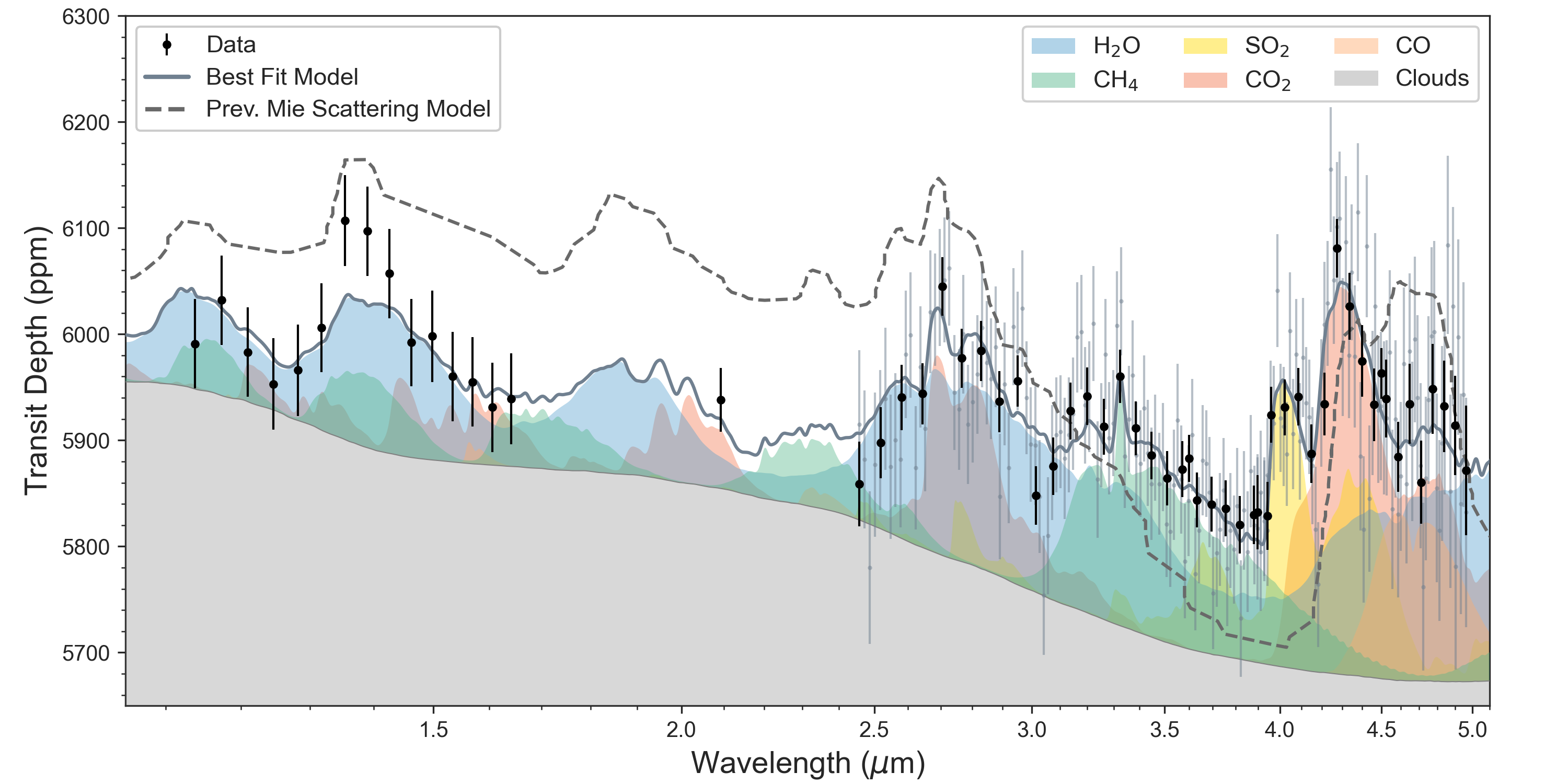}
\end{center}
\vskip -0.2in 
\caption{The combined NIRCam and HST/WFC3 transmission spectrum of GJ 3470 b showing major absorption features. The measured transit depths are plotted with their associated $1\,\sigma$ uncertainties. The NIRCam data have been binned into 0.06\um-wide spectral channels (black points) for visual clarity. The background grey points show the NIRCam data binned in 0.015\um-wide channels, which is the direct result of our analysis and was used in our atmospheric modeling. The solid dark-grey line shows the best fit model transmission spectrum from the atmospheric free retrieval analysis. The colored regions illustrate the contributions of various molecules and clouds to the best fit model. Note that the new NIRCam data and the refined HST transit depths strongly disfavor the previously suggested Mie scattering feature near 3\,\um\ \citep[dashed grey line,][]{benneke2019gj3470}.}
\label{fig:contrib}  
\end{figure*}

This stellar SED fitting (Figure \ref{fig:sed}) allowed us to measure the radius of the star GJ 3470 to be $R_*=0.50\pm0.01$\,\rsun. We then used the mean value of $R_P/R_*$ as measured in our joint broadband transit fits to estimate the planetary radius of GJ 3470 b to be $R_p=4.22\pm0.09$\,\re. Using the stellar density we measure as a part of the joint broadband transit fits, we constrain GJ 3470's stellar mass to be $M_*=0.44\pm0.04$\,\msun, and using a mass-ratio measured via radial velocity observations \citep{kosiarek2019}, we estimate a planetary mass of $M_p=11.1\pm0.9$\,\me. Our atmospheric retrievals use these absolute stellar and planetary radii and masses. 

\subsection{Stellar Variability Monitoring}

Previous photometric monitoring of GJ 3470 during the 2012-2013 season showed that the star itself is slightly variable at the presumed stellar rotation period of 21 days \citep{biddle2014}. This same monitoring campaign now spans from 2012 to 2023 and has given us a more precise measurement of stellar variability (Figure \ref{fig:starvar}). 

Using these new additional observations, we find that the GJ 3470 system shows sinusoidal brightness modulations with an amplitude of $2545\pm130$\,ppm at a period of $P=21.6239\pm0.0014$\,days. This variability amplitude is roughly half the value measured previously using only the 2012-2013 data \citep{biddle2014} and roughly half the variability amplitude used in previous modeling of how stellar variability could affect GJ 3470 b's transmission spectrum \citep{benneke2019gj3470}.

Previous work \citep{benneke2019gj3470} modeled the wavelength dependence of GJ 3470's variability and its effect on transmission spectra using a variability amplitude approximately twice that which we measured and found that it was at most 20\,ppm in the WFC3 bandpass and $<$10\,ppm over the F322W2 and F444W wavelength range. Since even our conservative estimate for the effect of stellar variability -- assuming a wavelength-independent amplitude -- is on the order of (or smaller) than our estimated depth uncertainties and the inter-instrument offsets, we do not expect the variability seen in the GJ 3470 system to affect our results significantly.  

\begin{figure*}[t!] 
\begin{center}                                 
\includegraphics[width=1.0\textwidth]{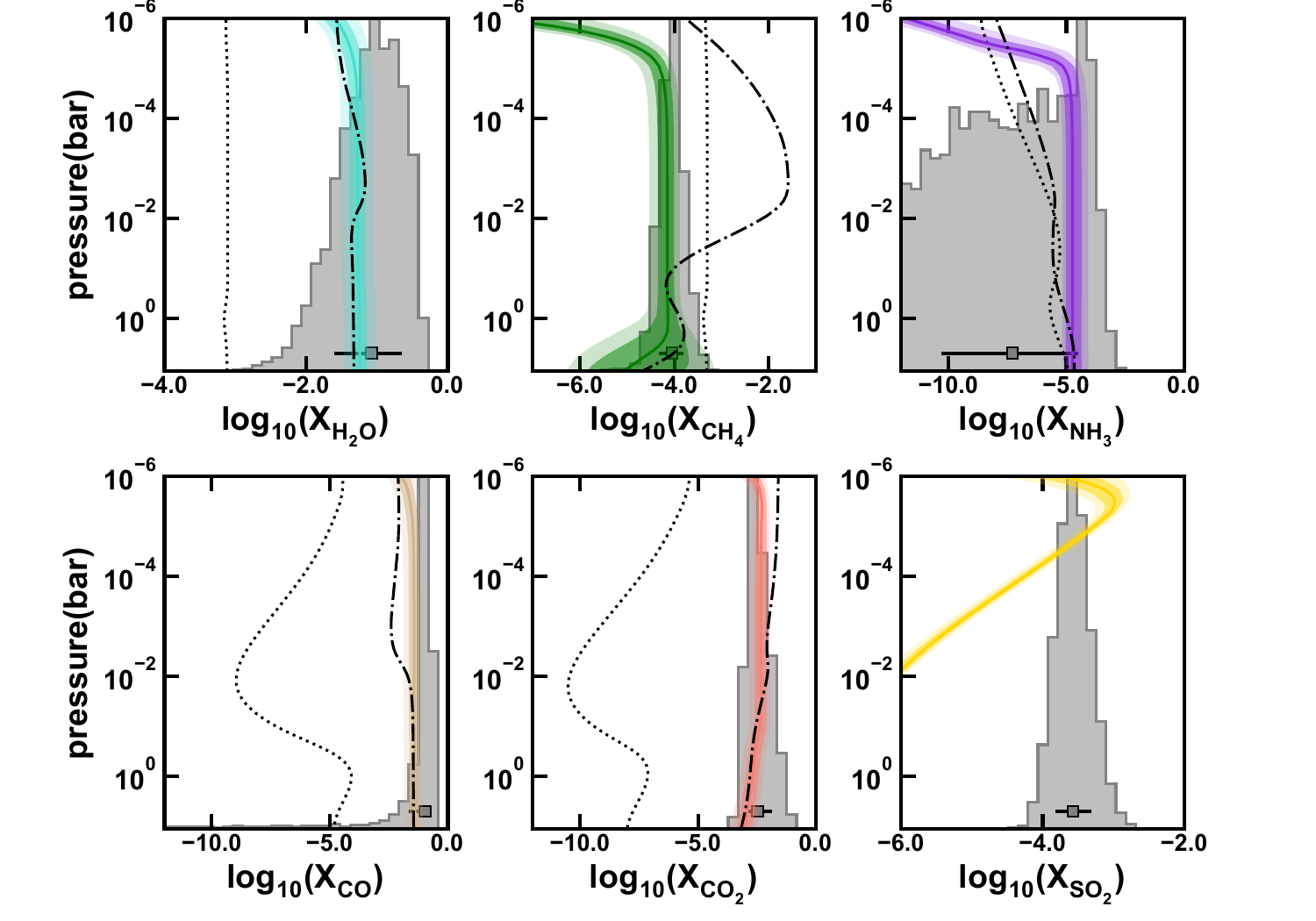}
\end{center}
\vskip -0.2in 
\caption{A comparison of the results from the 1D-RCPE and the free-retrieval atmospheric modeling, demonstrating their agreement. The free-retrieval's posterior distributions (gray) are shown against the abundance profiles (colored lines) from the 1D-RCPE grid modeling. The dotted lines show the expected composition of a Solar metallicity and Solar C/O atmosphere. The dash-dotted lines show the expected abundance profiles for a 125$\times$ solar metallicity atmosphere with a C/O of 0.35, as estimated for GJ 3470 b, under chemical equilibrium. The solid colored lines and shaded regions show the median 1$\sigma$ and $2\sigma$ profiles for GJ 3470 b under the disequilibrium conditions in the 1D-RCPE modeling. Note that for all molecules except for \methane\ and \sulfurdioxide\ the abundances estimated from the 1D and free-retrieval modeling match the equilibrium abundances of the 125$\times$ solar metallicity atmosphere with a C/O of 0.35. The indicates that \methane\ is likely being depleted and \sulfurdioxide\ enhanced via disequilibrium processes.}
\label{fig:abundances}  
\end{figure*}

\begin{figure*}[t!] 
\begin{center}                                 
\includegraphics[width=1.0\textwidth]{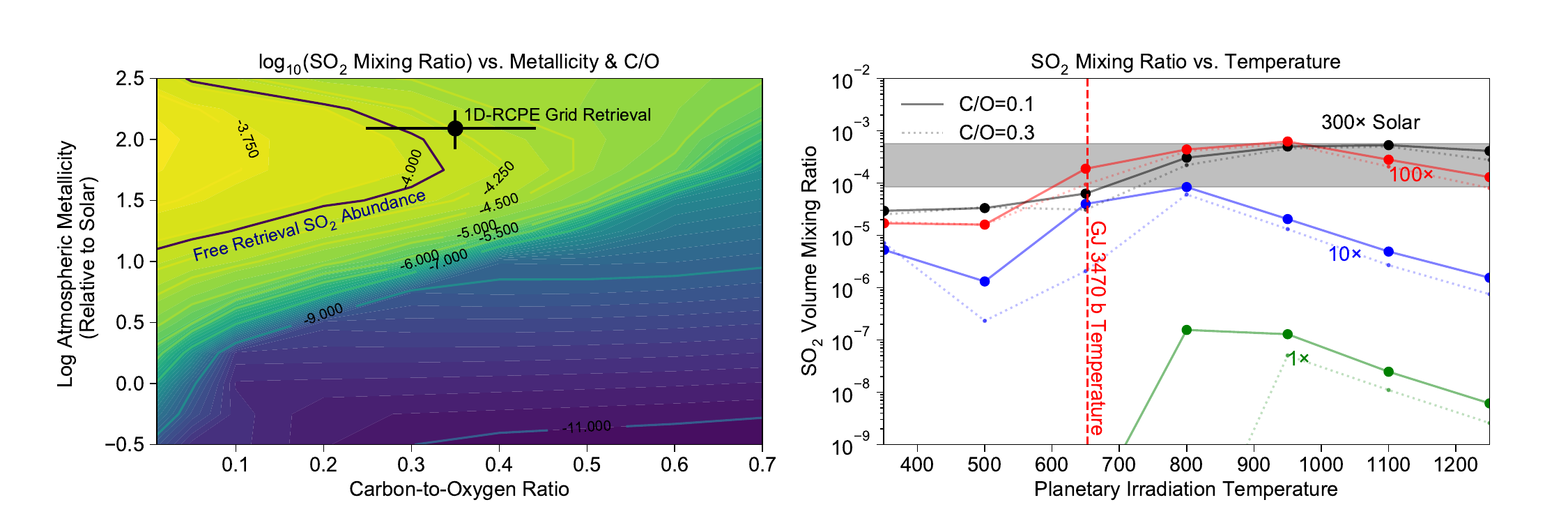}
\end{center}
\vskip -0.2in 
\caption{The impact of composition and temperature on the SO$_2$ abundance in cool ($<$800 K) sub-Neptune atmospheres. The left panel illustrates the column-averaged (1 $\mu$bar - 0.1 mbar) volume mixing ratio of SO$_2$ (contours) as a function of the atmospheric metallicity (relative to Solar) and the carbon-to-oxygen ratio for the nominal temperature of GJ 3470 b. There is a steep fall off in SO$_2$ with increasing C/O and increasing carbon-to-oxygen ratio. The right panel shows the dependency of the column-averaged SO$_2$ abundances with planetary irradiation temperature for representative metal enrichment. The presence of SO$_2$ at these temperatures substantially restricts the possible atmospheric composition. The retrieved SO$_2$ abundance (navy contour on the left and the shaded box on the right) strongly indicates an atmosphere with $\sim$100$\times$ stellar metallicity and a sub-stellar carbon-to-oxygen ratio.}
\label{fig:so2}  
\end{figure*}

\section{Atmospheric Modeling}

We interpreted the transmission spectra of GJ~3470~b using one-dimensional atmospheric models spanning a wide range of physico-chemical assumptions. First, we use parametric models coupled with Bayesian parameter estimations using nested sampling \citep{Buchner2014} (e.g., `free' atmospheric retrievals, see Methods) to quantify the detection of the multiple molecular features present in the spectrum of the planet. This paradigm independently fits for the chemical abundances, vertical pressure-temperature structure, and clouds in the planetary atmosphere without any assumptions of radiative-convective thermo-chemical equilibrium. 

The Bayesian model comparison results in strong detections of H$_2$O (6.3 $\sigma$),CO$_2$ (7.3 $\sigma$), CH$_4$ (3.8 $\sigma$), and SO$_2$ (4.0 $\sigma$), confirming the importance of the spectral features visible in the planetary spectrum (Figure \ref{fig:contrib}). Although the retrieved molecular abundances can be sensitive to the presence of possible offsets between the different instruments in GJ 3470 b's spectrum, the reported detections are robust regardless of any instrumental shifts and do not depend on the 163 ppm offset between NIRCAM F322W2 and F444W. The retrieved atmospheric abundances are suggestive of a super-solar metallicity with estimates of $\gtrsim100\times$~solar (e.g., $\log_{10}(X_{\textnormal{H}_2\textnormal{O}}) = -1.08 ^{+ 0.43 }_{- 0.52 }$, $\log_{10}(X_{\textnormal{CH}_4}) = -4.05 ^{+ 0.25 }_{- 0.27 } $, $\log_{10}(X_{\textnormal{CO}_2})= -2.47 ^{+ 0.61 }_{- 0.43 } $, and $\log_{10}(X_{\textnormal{SO}_2}) = -3.57 ^{+ 0.26 }_{- 0.25 }$). 

Then, to obtain a self-consistent estimate of the atmospheric metallicity and carbon-to-oxygen ratio we use 1-dimensional radiative-convective-photochemical equilibrium models \citep[1D-RCPE,][]{bell2023_methane}, which we describe in more detail in Appendix E. Briefly, the 1D-RCPE models self-consistently compute the atmospheric pressure-temperature profiles and gas volume mixing profiles using given values for the incident stellar flux, internal temperature, elemental abundances (via a scaled metallicity and C/O), and the vertical mixing strength. Using the same Bayesian inference methodology we derive the temperature, gas, and molecular weight profiles derived from the pre-computed 1D-RCPE models above are interpolated and post-processed via a transmission geometry radiative transfer routine (see Methods). As with the `free' retrieval the atmospheric models account for the effects of clouds and hazes using parameterizations for their inhomogeneous limb cover (see Methods).

The 1D-RCPE models agree with the free-retrieval interpretation of a super-solar atmospheric metallicity for GJ 3470 b, retrieving a metallicity of $125\pm40$ times Solar \citep[$\mathrm{M}/\mathrm{H}=2.1\pm0.15$ dex,][]{Lodders2009}, and a slightly sub-Solar carbon-to-oxygen ratio of $\mathrm{C}/\mathrm{O}=0.35\pm0.1$. The inferred molecular profiles (e.g., molecular volume mixing ratios as a function of height) are generally consistent with the inferred volume mixing ratios from the constant with height free-retrieval as shown in Figure \ref{fig:abundances}. The general agreement in the atmospheric composition between the flexible parametric free-retrieval and the 1D-RCPE retrieval confirms the interpretation of a super-solar metallicity atmosphere affected by disequilibrium processes (e.g., photochemistry) that give rise to SO$_2$ production. As shown in Figure \ref{fig:contrib}, this atmosphere model adequately explains the observed panchromatic spectrum, capturing nearly all the salient features.

\section{Discussion}

The detection of \sulfurdioxide\ in the atmosphere of GJ 3470 b extends the regime for sulfur-based photochemistry to lower planetary masses and cooler temperatures than previously probed. \sulfurdioxide\ has been detected in the atmospheres of WASP-39 b \citep[90\,\me, $\mathrm{T}_\mathrm{eq}=1166$\,K,][]{WASP39G395H,WASP39prism} and WASP-107 b \citep[30\,\me, $\mathrm{T}_\mathrm{eq}=770$\,K,][]{MIRI_EC_2024_W107}. GJ 3470 b is significantly lower in mass (11\,\me) and at a significantly lower equilibrium temperature ($\mathrm{T}_\mathrm{eq}=600$\,K\cite{bonfils2012}) than either of these planets. The detection of \sulfurdioxide\ in GJ 3470 b's atmosphere opens the door to using sulfur abundances to recover the formation history of sub-Neptunes \citep{turrini2021so2formation,crossfield2023so2formation}.

\begin{table}
    \centering
    \caption{Free-retrieval Atmospheric Properties.}
    \footnotesize
    \begin{tabular}{cl|ccc}
    \hline
     & Parameter  & Value  & Prior & DS  \\
    \hline
    \multirow{7}{*}{\rotatebox[origin=c]{90}{Chemical Species}}
    &$\log_{10} \left(X_{\textnormal{H}_2\textnormal{O}}\right)$&$-1.08 ^{+ 0.43 }_{- 0.52 }$&($-12.0,-0.3$) & 6.3$\sigma$ \\
    &$\log_{10}\left(X_{\textnormal{CH}_4}\right)$&$-4.05 ^{+ 0.25 }_{- 0.27 }$ &($-12.0,-0.3$) & 3.8$\sigma$ \\
    &$\log_{10}\left(X_{\textnormal{NH}_3}\right)$ &$-7.30 ^{+ 2.78 }_{- 3.02 }$ &($-12.0,-0.3$) & N/A \\
    &$\log_{10}\left(X_{\textnormal{HCN}}\right)$ &$-8.14 ^{+ 2.55 }_{- 2.46 }$ &($-12.0,-0.3$) & 1.2$\sigma$ \\
    &$\log_{10}\left(X_{\textnormal{CO}}\right)$&$-0.96 ^{+ 0.18 }_{- 0.70 }$ &($-12.0,-0.3$) & 1.5$\sigma$ \\
    &$\log_{10}\left(X_{\textnormal{CO}_2}\right)$&$-2.47 ^{+ 0.61 }_{- 0.43 }$&($-12.0,-0.3$) & 7.3$\sigma$\\
    &$\log_{10}\left(X_{\textnormal{SO}_2}\right)$& $-3.57 ^{+ 0.26 }_{- 0.25 }$&($-12.0,-0.3$) & 4.0$\sigma$\\
    \hline
    \multirow{6}{*}{\rotatebox[origin=c]{90}{P-T}} 
    &T$_{0}$ (K) &$412.54 ^{+ 67.65 }_{- 65.57 } $ & (300, 800) \\
    & $\alpha_1$  (K$^{-1/2}$) &$1.37 ^{+ 0.38 }_{- 0.37 } $  & (0.02, 2.0) \\
    &$\alpha_2$  (K$^{-1/2}$)  &$1.11 ^{+ 0.56 }_{- 0.55 } $  & (0.02, 2.0)\\
    &$\log_{10}(\mathrm{P}_1$) (bar) &$-2.54 ^{+ 2.05 }_{- 2.39 }$ & (-9.0, 2.0) \\
    &$\log_{10}(\mathrm{P}_2$)(bar) &$-6.11 ^{+ 2.52 }_{- 1.89 }$ & (-9.0, 2.0)\\
    &$\log_{10}(\mathrm{P}_3$) (bar) &$0.43 ^{+ 1.07 }_{- 1.34 }$ & (-2.0, 2.0) \\
    \hline
    &$\log_{10}(\mathrm{P}_\mathrm{ref.}$)  (bar)&$-5.24 ^{+ 2.95 }_{- 2.36 } $ & (-9.0, 2.0)\\
    &$\mathrm{R}_\mathrm{p}$  ($\mathrm{R}_\mathrm{Jup.}$ )&$ 0.38 ^{+ 0.01 }_{- 0.01 } $ & (0.30, 0.45)\\
    \hline
    \multirow{4}{*}{\rotatebox[origin=c]{90}{Cloud/Haze}} 
    &$\log_{10}(a)$ &$8.89 ^{+ 0.78 }_{- 1.16 }$ & (-4, 10)\\
    &$\gamma$ &$-8.10 ^{+ 1.32 }_{- 1.00 }$ & (-20, 2) \\
    &$\log_{10}(\mathrm{P}_\mathrm{cloud})$ (bar) &$-0.55 ^{+ 1.64 }_{- 1.59 }$ &(-9.0, 2.0) \\
    &$\phi_\mathrm{clouds and hazes}$ &$0.59 ^{+ 0.11 }_{- 0.07 }$&  (0, 1)\\
    \hline
    \\
    \end{tabular}
    \label{tab:free-retrieved}
\end{table}

Current models of sulfur photochemistry \citep{tsai2021vulcan,polman2023so2,tsai2023so2} indicate that at GJ 3470 b's temperature and mass, the \sulfurdioxide\ abundance should be low and more difficult to detect. This is predicted to occur because the oxidation reactions necessary to create \sulfurdioxide\ from \hydrogensulfide\ and OH would proceed over an order-of-magnitude more slowly than in the atmosphere of WASP-39b at the same metallicity and C/O due to the lower temperature on GJ 3470 b \citep{zahnle2016so2chem,tsai2023so2}. Note, however, that the existing literature on sulfur photochemistry predominantly focuses on Jupiter-like planetary atmospheres with metallicities of $\sim$1--10$\times$ Solar.

The presence of detectable \sulfurdioxide\ in GJ 3470 b's atmosphere likely indicates that the planetary atmosphere has a much higher metallicity, resulting in abundant water vapor. Models of sulfur photochemistry in \cite{tsai2023so2} predict that if the planetary atmosphere has a large amount of water vapor -- consistent with the high \water\ abundance we measure -- this will allow the UV-driven photolysis of \water\ on the planetary dayside to generate large amounts of OH radicals and overcome the slower chemical reaction rates due to the lower temperature. As shown in Figure \ref{fig:so2}, our modeling shows that to generate the \sulfurdioxide\ abundance we observe in GJ 3470 b ($\log_{10}\left(X_{\textnormal{SO}_2}\right) = -3.55^{+0.30}_{-0.52}$) the atmosphere needs to have a metallicity of $\sim$10--300$\times$ Solar and a C/O ratio $\lesssim0.35$, consistent with other parts of our analysis.

The \water\ and \carbondioxide\ abundances we measure in GJ 3470 b's atmosphere contrast with the low inferred \methane\ abundance of $\log_{10}\left(X_{\textnormal{CH}_4}\right) = -4.05 ^{+ 0.25 }_{- 0.27 }$. This is significantly higher than the $3\,\sigma$ upper-limit on the methane abundance of $\log_{10}\left(X_{\textnormal{CH}_4}\right) < -4.9$ reported in \cite{benneke2019gj3470}. At the same time, our measured \methane\ abundance is approximately 10$\times$ to 100$\times$ lower than one would expect from thermochemical equilibrium calculations at the inferred metallicity and C/O of the atmosphere (Figure \ref{fig:abundances}). UV photolysis is unlikely to cause the relatively low \methane\ depletion at the pressures probed by these data \citep{kempton2012gj1214,moses2014chemkinetics}. It is possible that the \methane\ level in GJ 3470 b's atmosphere is being reduced by vertical chemical quenching and the transport of \methane-poor and CO-rich gas up to the pressures probed by these observations \cite{venot2014gj3470}. \methane\ depletion via vertical disequilibrium processes has been suggested for other exoplanet atmospheres recently observed with JWST \citep{WASP39prism,xue2024hd209458}. In GJ 3470b, the \methane\ abundance may be further reduced due to tidal heating. Spitzer eclipse observations show that GJ 3470 b's orbital eccentricity is significantly non-zero, with $e\cos\omega=0.01457\pm0.0075$ \citep{benneke2019gj3470}, consistent with the RV observations \citep[e.g.,][]{kosiarek2019,Gumi2022PolarOrbit}. For a Neptune-like tidal quality factor of $Q = 10^4$, the planet's internal temperature would be T$_{\mathrm{int}}\approx550$\,K. This would increase vertical mixing and decrease the \methane\ abundance to match our observations \citep{Fortney2020Tint}. Note that our modeling only provides a loose constraint of T$_{\mathrm{int}}=250\pm200$\,K.

The strong signature of \sulfurdioxide\ in our data supports the interpretation that GJ 3470 b possesses a high metallicity, near Solar C/O, atmosphere with reduced \methane\ from disequilibrium processes. In principle, an analysis using atmosphere models assuming thermochemical equilibrium might find a consistent solution that accounts for the \water, \carbondioxide, and \methane\ abundances we measure -- likely a Solar metallicity and very low C/O atmosphere as previously reported for GJ 3470 b in \cite{benneke2019gj3470} However, the significant \sulfurdioxide\ we measure in GJ 3470 b is inconsistent with equilibrium models \citep{tsai2021vulcan,polman2023so2}. As discussed above, the presence of \sulfurdioxide\ is itself indicative of disequilibrium processes and a high metallicity atmosphere (see also Figure \ref{fig:so2}). Thus, we consider this the preferred scenario.

Recently, \cite{turrini2021so2formation} and \cite{crossfield2023so2formation} have proposed using the combination of carbon, oxygen, and sulfur abundances in exoplanet atmospheres to estimate their initial formation locations in the protoplanetary disk, and possibly the formation process itself \citep[i.e. planetesimal vs. pebble accretion,][]{crossfield2023so2formation}. Specifically, both papers suggest that the $\mathrm{C}/\mathrm{S}$ and $\mathrm{O}/\mathrm{S}$ ratios in a planetary atmosphere should increase by a factor of 3x-20x relative to Solar as the initial formation location increases from 5-30\,AU in the case of pebble accretion. The atmospheric abundances that we measure in GJ 3470 b yield rough estimates of $\mathrm{C}/\mathrm{S} = 15\pm10$ times Solar and $\mathrm{O}/\mathrm{S} = 20\pm10$ times Solar. This suggests pebble accretion at 10-30\,AU from the host star, but better constraints are needed on the CO abundance and the $\mathrm{H}_2\mathrm{S}$ abundance to say anything definite.

The results from these observations further illustrate the new complexity in exoplanet atmospheres that is becoming visible with JWST observations. The observed \methane\ depletion and strong signature of \sulfurdioxide\ indicate that GJ 3470 b's atmosphere is a dynamic system -- both kinetically and chemically -- that is not easily summarized without an accounting of all the major carbon- and oxygen-bearing molecules. Though the high metallicity and Solar C/O we infer for GJ 3470 b are both more in line with the theoretical expectations from planet formation models, the compositional diversity being revealed by JWST data indicates that relating observations of atmospheric compositions to formation histories may be a more idiosyncratic process than has been expected. The clear detection of \sulfurdioxide\ on a cool planet like GJ 3470 b also indicates that photochemistry is an important consideration across a wide range of temperatures. Scheduled future observations of GJ 3470 b's transmission spectrum using JWST/NIRSpec G395H will hopefully further illustrate the intriguing complexity of this atmosphere and further constrain its formation history by providing an independent measurement of the atmospheric \sulfurdioxide\ abundance.

\begin{acknowledgments}
This work has made use of NASA's Astrophysics Data System, the Extrasolar Planet Encyclopedia at exoplanet.eu \citep{exoplanetseu}, the SIMBAD database operated at CDS, Strasbourg, France \citep{simbad}, and the VizieR catalog access tool, CDS, Strasbourg, France \citep{vizier}. This research has made use of the NASA Exoplanet Archive, which is operated by the California Institute of Technology, under contract with the National Aeronautics and Space Administration under the Exoplanet Exploration Program. L.W.~acknowledges support for this work provided by NASA through the NASA Hubble Fellowship grant HST-HF2-51496.001-A awarded by the Space Telescope Science Institute, which is operated by the Association of Universities for Research in Astronomy, Inc., for NASA, under contract NAS5-26555. T.J.B.~and T.P.G.~acknowledge funding support from the NASA Next Generation Space Telescope Flight Investigations program (now JWST) via WBS 411672.07.05.05.03.02. M.R.L.~acknowledges NASA XRP award 80NSSC19K0446 and STScI grant HST-AR-16139. M.R.L.~and L.W.~acknowledge Research Computing at Arizona State University for providing HPC and storage resources that have significantly contributed to the research results reported within this manuscript. K.O. acknowledges support from JSPS KAKENHI Grant Number JP23K19072. M.M. acknowledges funding from NASA Goddard Spaceflight Center via NASA contract NAS5-02105. G.W.H. acknowledges long-term support for our automatic telescopes from NASA, NSF, and Tennessee State University. We thank Marcia Rieke for allocating the NIRCam time for this program.
\end{acknowledgments}


\vspace{5mm}
\facilities{HST (WFC3), JWST (NIRCam), Spitzer (IRAC)}


%\software{astropy \citep{2013A&A...558A..33A,2018AJ....156..123A},  
%          Cloudy \citep{2013RMxAA..49..137F}, 
%          Source Extractor \citep{1996A&AS..117..393B}
%          }



\clearpage
\newpage

\appendix

\section{Data Reductions and Comparisons}

\subsection{JWST Data Reduction}

As mentioned, to ensure the reproducibility of our results we performed three separate analyses of the NIRCam spectra using the \texttt{Pegasus}, \texttt{Eureka!} \citep{bell2022eureka}, and \texttt{tshirt} \citep{ahrer2023wasp39,bell2023_methane} data reduction pipelines (Figure \ref{fig:spectrum}). We reduced both of the LW spectroscopic transit observations using three separate data reduction pipelines, which we then compared as a self-consistency check. We performed a single reduction on the SW photometric observations. 

\subsubsection{Pegasus Data Reduction}

Our fiducial NIRCam data reduction used the \texttt{Pegasus} pipeline (\url{https://github.com/TGBeatty/PegasusProject}), which was initially developed for the NIRCam commissioning observations. Our reduction began with background subtraction on the \textsc{rateints} files provided by version 1.10.2 of the \texttt{jwst} pipeline using CRDS version 11.17.0. For each \textsc{rateints} file, we first performed a general background subtraction step by fitting a two-dimensional, second-order spline to each integration using the entire 256$\times$2048 \textsc{rateints} images. To fit the background spline, we first masked out image rows 5 to 75 to not self-subtract light from the GJ 3470 system and from rows 130 to 140 in the F322W2 images to mask a nearby background star. We then performed a single round of $3\,\sigma$ clipping on the unmasked portions of the image. We then individually fit a spline to each amplifier region on the image using the unmasked, unclipped pixel values with a median-box size of 20 pixels. Fitting the background spline on a per-amplifier basis is necessary to remove residual bias offsets between the amplifier regions. We extrapolated the combined background spline for the whole image over the masked portions near GJ 3470 and subtracted it from the original image values. Visual inspection of the \textsc{rateints} images showed that in roughly 5\% of the integrations the reference pixel correction failed for at least one of the amplifier regions, so after the spline fitting and subtraction we re-ran the reference pixel correction using \texttt{hxrg-ref-pixel} (\url{https://github.com/JarronL/hxrg_ref_pixels}). Finally, we attempted to remove some of the red-noise caused by the NIRCam readout electronics along detector rows, by calculating the biweight mean of each row using pixels from column 1800 onwards for the F322W2 images and up to column 600 for F444W (both chosen to avoid light from GJ 3470), and then subtracting this mean from each row. Visually, this removed most of the horizontal banding typical for NIRCam \textsc{grismr} images.

We then extracted broadband and spectroscopic lightcurves from our background-subtracted images. To do so, we fit the spectral trace using a fourth-order polynomial and then used optimal extraction to measure the 1D spectrum in each image. We performed three rounds of iterative profile estimation for the optimal extraction routine, after which we judged the profile fit to have converged. Using the resultant 1D spectra, we extracted a broadband lightcurve from 2.45\um\ to 3.95\um\ at F322W2 and from 3.89\um\ to 4.97\um\ at F444W. For the spectroscopic lightcurves, we subdivided each of these wavelength regions into individual 0.015\um-wide spectral channels. For both the broadband and spectroscopic lightcurves, we linearly interpolated over each spectral column to account for partial-pixel effects in the wavelength solutions.

\subsubsection{Eureka! Data Reduction}

We also reduced the observations using the open-source \texttt{Eureka!} pipeline \citep{bell2022eureka}. Our Eureka! reduction of the F322W2 data followed the same method as described by \cite{bell2023_methane}, with the only difference being the package versions used. Our \texttt{Eureka!} reduction used version 0.10.dev15+g80126b56.d20230613 of the \texttt{Eureka!} pipeline, CRDS version 11.17.0 and context 1093, and \texttt{jwst} package version 1.10.2. The \texttt{Eureka!} Control Files and Parameter Files we used are available for download (\url{https://doi.org/10.5281/zenodo.10779291}). In summary, we cropped the full subarray images only to use y-pixels 5--64 and x-pixels 15--1709 and then corrected for the curvature of the spectral trace. We then performed per-column, per-integration background subtraction on the Stage 2 outputs; and performed optimal spectral extraction using the pixels within 9 pixels of the center of the flattened spectral trace. For each spectroscopic lightcurve, we then masked 10-sigma outliers compared to a boxcar-filtered version of the lightcurve (with a boxcar width of 50 integrations) to remove any remaining cosmic rays while not removing sharp features like transit ingress/egress.

Our \texttt{Eureka!} reduction of the F444W data used the same package versions as the F322W2 observations. We ran Stages 1 and 2 of \texttt{Eureka!} using the same parameters as for the F322W2 reduction. For Stage 3, we cropped each image to only include x-pixels 738--2000 (since the spectra fall on a different region of the detector), and we used a smaller aperture half-width of 4 when extracting the spectra. 

\subsubsection{tshirt Data Reduction}

Finally, we also similarly reduced the data as for WASP-80 b \citep{bell2023_methane} using the \texttt{tshirt} \citep{ahrer2023wasp39,bell2023_methane} pipeline. We used data from SDP\_VER = 2022\_3b, JWST version 1.6.0, CRDS version 11.16.5 for the stage 1 (\texttt{calwebb\_detector1}) processing, with some modifications. The CRDS context was jwst\_1011.pmap for the F322W2 filter and jwst\_1009.pmap for the F444W filter, but the reference files used were the same. The modifications to the \texttt{calwebb\_detector1} were to replace the reference pixel step with a Row-by-row Odd/Even Subtraction (ROEBA) step using sky pixels\cite{schlawin2023}. We also skip the \texttt{dark\_current} step as the subarray dark subtraction adds noise to the time series, and finally we adjust the jump step rejection threshold to be 6$\sigma$ to avoid spurious false positive jump step detections. At the completion of \texttt{calwebb\_detector1}, we divide the NRCALONG detector image by the imaging flat field in the corresponding band.

For the spectroscopic extraction, we subtracted the background column-by-column with a linear robust fit and summed over a rectangular aperture with a covariance-weighted extraction profile \citep{schlawin2020}. The background linear fit was from pixels that were 10 to 30 pixels from the source centroid. The aperture was a 10 pixel wide aperture centered on pixel Y=34 for the F322W2 filter and Y=31 for the F444W filter. We assumed a read noise correlation of 0.08 between pixels and a read noise of 14 e$^-$.

\subsubsection{Shortwave F210M NIRCam Photometry}

NIRCam can collect shortwave photometric observations simultaneously with longwave spectroscopic observations. To reduce both of the photometric F210M observations taken along with the spectroscopic F322W2 and F444W transit observations, we used the \texttt{Eureka!} data analysis pipeline version 0.9, \citep{bell2022eureka} with CRDS version 11.16.20, \texttt{jwst} version 1.6.2. The F322W2 reduction used the CRDS context 1041 and the F444W reduction used 1046. We began by running Stages 1 to 4 of \texttt{Eureka!} on the uncalibrated, \textsc{uncal}, image files. For Stage 1, we used the default \texttt{jwst} settings, except that we changed the cosmic ray jump rejection threshold to 10$\sigma$ to avoid excessive false positive cosmic ray rejections. We ran Stage 2 with default settings, except that we did not perform the photometric calibration step.

To run Stage 3 on the F210M data, we first cropped the images to focus on the source location on the detector. For the photometry collected along with the F322W2 transit, we cropped to y-pixels 0--255 and x-pixels 550--1450, and y-pixels 20--250 and x-pixels 1000--2040 for the F444W transit. We then used a cubic interpolation method to fill in bad pixels and outliers in these cropped images. We identified bad pixels using the data quality array, and 7$\sigma$ outlier pixels along the time axis. We then used \texttt{Eureka!}'s version of the ROEBA \citep{schlawin2023} background correction technique to remove 1/$f$ noise in the images. To do so, we excluded pixels within 350 pixels of the location of GJ 3470 on the detector. We also measured the centroid of GJ 3470's point-spread function in each integration in a two-step process. We first estimated the center-of-light over an entire cropped image to get an initial guess at the centroid location. We then performed another center-of-light estimate using just an 11-pixel by 11-pixel box centered on the initial centroid guess. 

We extracted lightcurves using the resultant calibrated images from both F210M photometric observations using simple aperture photometry. We used a circular extraction aperture with a radius of 65 pixels for both observations, and a background annulus with inner and outer radii of 70 and 90 pixels for F322W2's shortwave observations and 75 and 90 pixels for F444W's shortwave observations. Finally, in \texttt{Eureka!}'s Stage 4, we clipped 8$\sigma$ outliers compared to a boxcar smoothed version of the data using a boxcar width of 50 integrations. The resulting F210M lightcurves contained 1681 points for the F322W2 transit with a median uncertainty of $\pm234$\,ppm per point, and 1009 points for the F444W transit with a median uncertainty of $\pm170$\,ppm per point.

\subsection{HST Data Reduction}

We reduced the spectroscopic observations and extracted broadband and spectroscopic lightcurves using the pipeline described in \cite{beatty2017kepler13}. In brief, we extracted 1D spectra from each of the scanned, up-the-ramp readouts using the G141 disperser after performing a set of bad-pixel and scan-angle corrections. We summed the 1D spectra from each of these subexposures together to make a single 1D spectrum for each exposure. Next, we extracted a broadband lightcurve and a set of spectroscopic lightcurves using the summed 1D spectra. The broadband lightcurve covered the wavelength range from 1.12\um\ to 1.66\um\ in a single channel, while the spectroscopic lightcurves covered this same wavelength range in fifteen bins that were each 0.036\um\ wide. This wavelength range and the spacing match the previous analysis of these data \citep{benneke2019gj3470}. The broadband and spectroscopic HST/WFC3 light curves for each visit contained 72 points split evenly between the visit's four HST orbits. The median flux uncertainty for the broadband lightcurve was 100 ppm, and the median flux uncertainty among all the spectroscopic lightcurves was 286 ppm.

\begin{figure*}[t!] 
\begin{center}                                 
\includegraphics[width=1.0\textwidth]{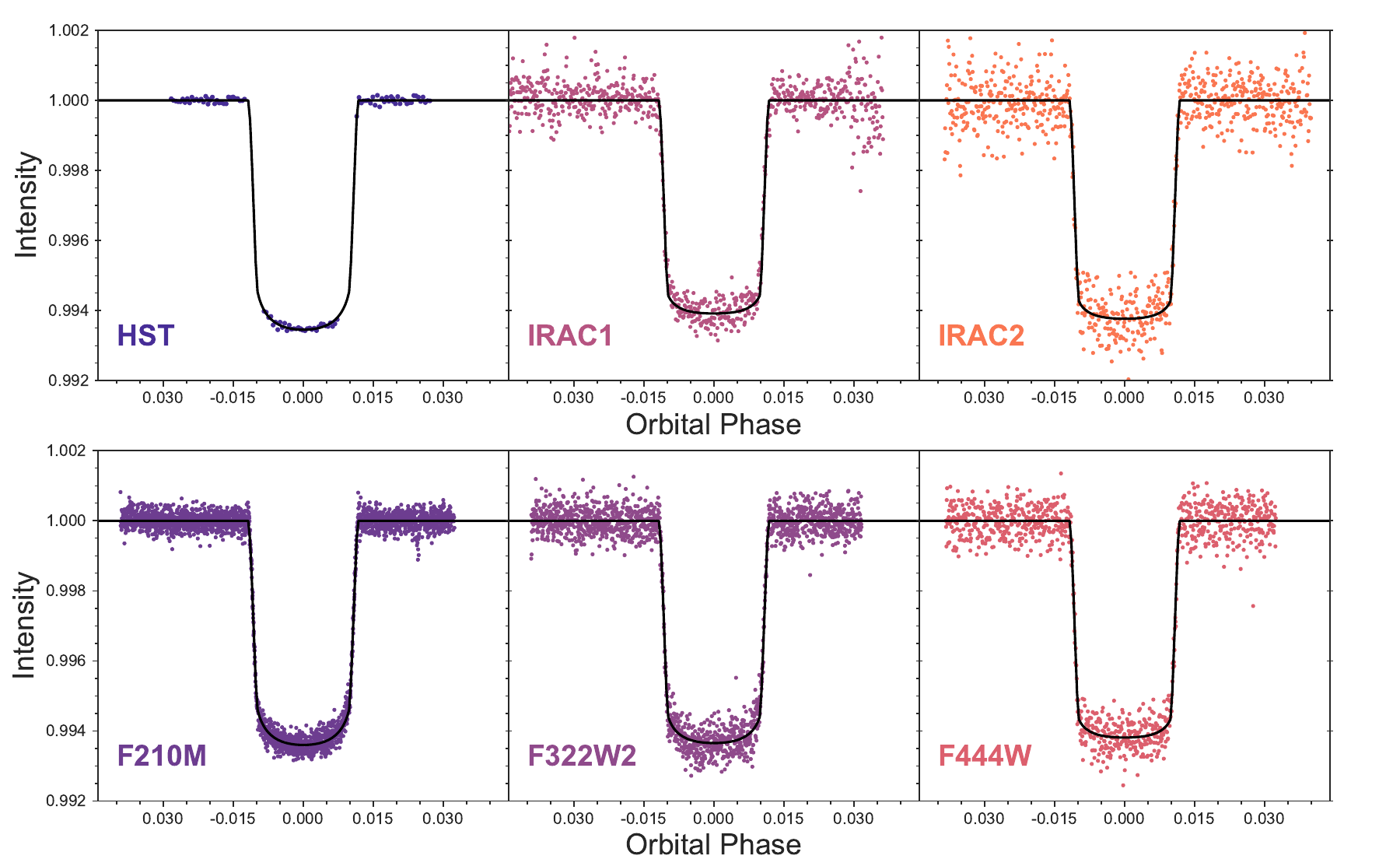}
\end{center}
\vskip -0.2in 
\caption{The broadband lightcurve data and bestfit transit models used in this analysis. To refine the orbital parameters of GJ 3470 b, we simultaneously fit archival broadband HST/WFC3 and Spitzer/IRAC data (top panels) together with the new JWST photometric (F210M) and spectroscopic (F322W2 and F444W) data (bottom panels). This provides significantly more precise measurements of the orbital ephemerides and the orbital separation. We use these results to fix the orbital properties of GJ 3470 b in our spectroscopic fitting. The full results of the joint broadband fit are given in Table 2. Note that the F210M panel shows both F210M lightcurves overplotted on top of each other: one from each of the F322W2 and F444W observations.}
\label{fig:broadband}  
\end{figure*}

\subsection{Spitzer Data Reduction}

We began our data reduction and photometric extraction process from the basic calibrated data (BCD) images. We first subtracted the background from the images and corrected for bad pixels. We then used the background-subtracted, bad-pixel corrected images to measure the pixel position of GJ 3470 in each image using a two-dimensional Gaussian. Note that we used these corrected images only to estimate the background and to measure the position of GJ 3470 -- we used uncorrected background-subtracted images for the photometric extraction.

We extracted raw photometry for GJ 3470 using a circular extraction aperture centered on the star's position in each image. We used an aperture radius of 2.4 pixels. For reference, the average full-width half-maximum of GJ 3470's point spread function was 2.05 pixels at \three and 2.00 pixels at \fouralt. Since the fitting process for these observations was time-intensive, we did not perform a complete optimization to determine the best extraction aperture size. Instead, we used a fixed aperture radius of 2.4 pixels for all the datasets, matching the average optimum aperture size estimate previously in \cite{benneke2019gj3470}. We performed a limited test of our aperture size by extracting and fitting photometry for aperture radii of 2.2 and 2.6 pixels. In both cases, the log-likelihoods of the resulting best fits were lower, the scatter in the residuals higher, and the transit depths consistent with our optimum aperture at 2.4 pixels.

We trimmed the first 250 data points for each observation to remove residual ramp effects and performed a single round of $5\,\sigma$ clipping to remove outliers. The median uncertainty at \three was 4462 ppm for all the observations, and at \four was 6073 ppm.

\section{Lightcurve Fitting}

To measure the transmission spectrum of GJ 3470 b, we first performed a simultaneous, joint fit of all six of the broadband lightcurves (Figure \ref{fig:broadband}) and then used the resulting orbital properties for GJ 3470 b as fixed input to our spectroscopic fits to the WFC3, F322W2, and F444W data. Our intention in doing so, and in refitting transit depths that have been previously reported, was to construct a self-consistent transmission spectrum for GJ 3470 b across all of our data without possible offsets caused by different priors on the planetary orbital parameters and different limb-darkening properties. In particular, we measure the transit depth in the WFC3 data to be approximately 100 ppm shallower than in \cite{benneke2019gj3470} solely owing to more refined measurements of $a/R_*$ and $\cos(i)$ driven by the JWST data.

For the joint broadband fit, we used Gaussian priors on GJ 3470 b's $a/R_*$ \citep{awiphan2016}, orbital inclination \citep{awiphan2016}, $e\cos\omega$ \citep{benneke2019gj3470}, $e\sin\omega$ \citep{kosiarek2019}, transit center time \citep{awiphan2016}, and period \citep{awiphan2016}. We did not impose priors on the planet-to-star radius ratio in any of the observed bands.

In all of our broadband and spectroscopic fitting, we used limb-darkening coefficients calculated from the Kurucz ATLAS9 \textsc{odfnew} stellar models \citep{atlas9}. We did so using the nominal spectroscopic stellar properties for GJ 3470 \citep{kosiarek2019}.

\subsection{Joint Broadband Transit Fitting}

To measure the orbital and system properties of GJ 3470 b, we first fit broadband HST, broadband JWST, and Spitzer transit lightcurves simultaneously (Figure \ref{fig:broadband}). We fit a single set of orbital and system parameters for all seven lightcurves and individual values for $R_p/R_*$ for each data set and filter. The exceptions to this were the two F210M lightcurves, which we fit using a single shared value of $R_p/R_*$ for both. We used \texttt{BATMAN} \citep{batman} transit models and performed an MCMC fit to the broadband data using the \texttt{emcee} \citep{emcee} Python package. We imposed Gaussian priors on some of the transit parameters as described above. We judged the MCMC fitting to have converged ny checking the Gelman-Rubin statistic for each parameter was below 1.1, and by visual inspection of the posterior corner plot. We also checked the Gaussianity of the residuals for each of the broadband lightcurve fits using an Anderson-Darling test, and we did not find any significant non-Gaussianity.

We detrended all three of the JWST broadband datasets (F210M, F322W2, and F444W) using linear trends in time and a free normalization factor. We allowed for different slopes between each of the F210M visits. For the HST broadband lightcurve we detrended the usual systematics seen in WFC3 lightcurves using a combined linear ramp and exponential ramp -- with exactly the same functional form as we describe below in the fitting process for the WFC3 spectroscopic lightcurves.

The broadband Spitzer lightcurves showed position-dependent flux variations caused by intrapixel effects that are typical for these observations. We used a linear ramp coupled with a Bi-Linear Interpolated Subpixel Sensitivity (BLISS) map \citep{blissmap} to detrend the Spitzer lightcurves. Recall that there are a total of seven \three transits and eight \four transits of GJ 3470 b that have been observed by Spitzer. To fit all these data simultaneously, we constructed a single, combined, \three BLISS map from all seven of the \three transits, and a single, combined, \four BLISS map from all eight of the \four transits. This technique has been used previously \citep{beatty2019}, since the underlying intrapixel sensitivity variations in both \textsc{irac} channels appear to be stable over multi-year timescales \citep{beatty2019,murphy2023wasp43var}. We did allow for individual background linear slopes between all fifteen Spitzer transits.

The results from the joint broadband fit are listed in Extended Data Table 1. Note that, although we fit for broadband transit depths using the WFC3, F322W2, and F444W data, we did not use these broadband depths in our subsequent atmospheric modeling. 

\subsection{JWST Spectroscopic Transit Fitting}

Next, we performed spectroscopic transit to fits the F322W2 and F444W data. To do so, we fixed the transit center time, orbital period, scaled semi-major axis, orbital period, eccentricity, and the argument of periastron to the bestfit values from the joint broadband fit (Table \ref{table:broadband}). As with our analysis of the broadband data, we used quadratic limb-darkening coefficients estimated from the Kurucz ATLAS9 \textsc{odfnew} stellar models \citep{atlas9}. We calculated limb-darkening coefficients for each of the spectral channels, and fixed the limb-darkening coefficients in each channel to these model values using the stellar spectroscopic properties from \cite{awiphan2016}. We experimented with using a non-linear limb-darkening law and allowing the limb-darkening coefficients to float as freely fit parameters and found that this did not significantly affect our results.

Similar to our data reduction and extraction approach, we used three separate analysis pipelines to fit the F322W2 and F444W transmission spectra as a self-consistency check and a single analysis pipeline on the WFC3 data. Note that each of the three analysis pipelines we used for the JWST data fit lightcurves extracted from their corresponding data reduction routine described above. We tested the different transit fitting routines on the lightcurves produced by the other pipelines (e.g., \texttt{Pegasus} lightcurve fitting on \texttt{Eureka!} and \texttt{tshirt} timeseries data) and found no significant differences in the end result.

We experimented using linear, quadratic, and no temporal detrending terms in the F322W2 and F444W spectroscopic fits. We found that a linear temporal trend was strongly preferred by a Bayesian Information Criteria (BIC) analysis over both the quadratic ($\Delta$BIC\,$\geq6$) and no detrend ($\Delta$BIC\,$\geq8$) in all spectral channels in both filters. We, therefore, used a linear trend in time in all three of our transit fitting analyses.

The measured transmission spectra from our fiducial \texttt{Pegasus} reduction of the F322W2 and F444W data are given in Tables 3 and 4. Note that in Table 3, the listed depths have not been offset by the 163 ppm offset we estimated for the F322W2 data and discuss in Section 2.4 and Appendix B.3.

\subsubsection{Pegasus}

We used a \texttt{BATMAN} \citep{batman} transit model in our Pegasus lightcurve fitting. Most of the transit model parameters were fixed to the values determined by the joint broadband fit, which left the free parameters in our spectroscopic lightcurve fitting to be the planet-to-star radius ratio and the slope and normalization of the background linear trend. We did not impose a prior on any of these parameters. We fit each spectral channel individually. 

To fit the spectroscopic lightcurves we performed an initial likelihood maximization using a Nelder-Mead sampler followed by MCMC likelihood sampling. We used the maximum likelihood point identified by the Nelder-Mead maximization as the starting locus for initializing the MCMC chains. To perform the MCMC runs, we used the \texttt{emcee} \citep{emcee} Python package using 12 walkers with a 2,000-step burn-in and then a 4,000-step production run for each spectral channel. We checked that the MCMC had converged by verifying that the Gelman-Rubin statistic was below 1.1 for each parameter in each spectral channel.

We additionally checked the goodness-of-fit and statistical properties of our transit modeling in each spectral channel. We did so by first verifying that the average of the per-point flux uncertainties in each channel's lightcurve matched the standard deviation of the residuals to the bestfit transit model. We also computed the Anderson-Darling statistic for each channel's lightcurve residuals to check that the residuals themselves appeared Gaussian. We did not find statistically significant non-Gaussianity in the residuals to our spectroscopic lightcurve fits.

\subsubsection{Eureka!}

Our astrophysical model consisted of a \texttt{starry} \citep{starry} transit model on which we placed a broad, uninformative prior on the transit depth. We fixed our orbital parameters to those from the joint broadband fit. In addition to the linear trend in time, our systematic model included a linear correlation with the changes in the spatial position and spatial PSF-width. We also fitted a white-noise multiplier to ensure a reduced chi-squared of 1 and avoid artificially constrained posteriors. No ``mirror tilt'' events \citep{schlawin2023} were evident in either observation.

We sampled the posteriors using \texttt{PyMC3}'s No U-Turns Sampler \citep{pymc3} with two independent chains, each taking 4,000 tuning draws and 3,000 posterior samples with a target acceptance rate of 0.85. We confirmed that the fits had converged by ensuring the Gelman-Rubin statistic was below 1.1 and by visual inspection of the two chains. Our best-fit values and uncertainties were computed using the 16th, 50th, and 84th percentiles of the \texttt{PyMC3} samples. For F322W2, we find that the noise is generally $\sim$20--30\% above the photon noise limit, while F444W is $\sim$2-10\% above the photon noise limit. Visual examination of the Allan variance plots \citep{Allan1966} for both observations show no evidence for residual red noise.

\begin{figure*}[t!] 
\begin{center}                                 
\includegraphics[width=1.0\textwidth]{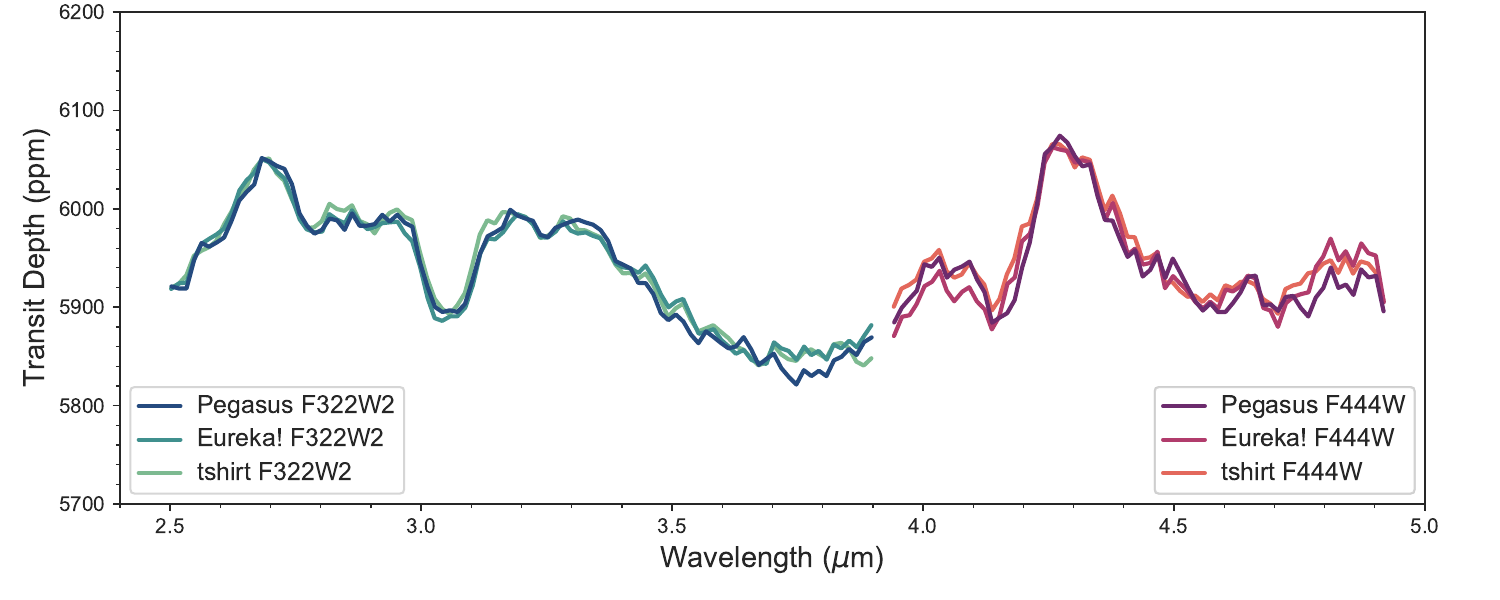}
\end{center}
\vskip -0.2in 
\caption{A comparison of the three different NIRCam data reductions. The NIRCam F322W2 and F444W transmission spectra from all three of the data reduction pipelines, smoothed using a 3-point wide boxcar filter for clarity. The F322W2 spectra have all been offset down by 163\,ppm (see discussion in the text). All three spectra show spectra features with similar shapes and amplitudes. Compared to our nominal \texttt{Pegaus} spectra, the \texttt{Eureka!} and \texttt{tshirt} results differ by $\chi^2/\mathrm{dof}=0.13$ for \texttt{Pegasus} vs. \texttt{Eureka!}, and $\chi^2/\mathrm{dof}=0.21$ for \texttt{Pegasus} vs. \texttt{tshirt}.}
\label{fig:redcomp}  
\end{figure*}

\subsubsection{tshirt}

For our \texttt{tshirt} fitting, we modeled the lightcurves using a \texttt{starry} transit model \citep{starry}. We included two linear background trends: one with respect to time and one with respect to the focal plane array housing (FPAH) temperature. For the FPAH temperature, we used the JWST engineering database to look up the \texttt{IGDP\_NRC\_A\_T\_LWFPAH1} telemetry at the time of the observations. Since the telemetry data points are collected at a different cadence than our observations, we fit a 5th-order polynomial to the raw temperature values and used this to interpolate the FPAH temperatures at the time of the observations in our data. For the \texttt{tshirt} fitting, we binned the lightcurve and corresponding FPAH temperatures down to 300 points (roughly a factor of 5 binding for the F322W2 data and a factor of 3 binning for the F444W data) for faster lightcurve sampling.

Similar to the other reductions, we fixed the orbital properties of GJ 3470 b to the values determined in the joint broadband fit. The \texttt{tshirt} fit included an additional noise-scaling parameter, which scaled the overall uncertainties on the flux measurements to allow for an increased uncertainty over the photon and read-noise expectations.

To fit the spectroscopic lightcurves, we first performed a likelihood maximization to determine an initial best fit. We used this initial solution to identify and discard flux outliers that were more than 5$~\sigma$ from the model. We then used the remaining points to perform a No-U Turns sampling routine via \texttt{pymc3} \citep{pymc3}. We sampled the lightcurves with a target acceptance of 0.90, 3,000 tuning samples and 3,000 draws from the posterior with two chains.

\subsubsection{Comparison Between the JWST Reductions}

The motivation for conducting three different reductions of the NIRCam data was to check the robustness of our final transmission spectrum against different choices and assumptions made during the image calibration, spectral extraction, and lightcurve fitting stages of our analyses. Generally, we find that all three reductions give the same general transmission spectrum with approximately the same depth uncertainties (Figure \ref{fig:spectrum} and \ref{fig:redcomp}). The mean depth uncertainty across the entire NIRCam wavelength range is approximately 60 ppm at our constant wavelength $\Delta \lambda = 0.015$\,$\mu$m binning, which is larger than the mean point-to-point difference between the \texttt{Pegasus} and \texttt{Eureka!} spectra (28 ppm) and the \texttt{Pegasus} and \texttt{tshirt} spectra (29 ppm). A $\chi^2$ comparison of the three spectra gives $\chi^2/\mathrm{dof}=0.13$ for \texttt{Pegasus} vs. \texttt{Eureka!}, and $\chi^2/\mathrm{dof}=0.21$ for \texttt{Pegasus} vs. \texttt{tshirt}. The per-point uncertainties on the F322W2 and F444W spectra from all three reductions were similar.

\subsection{Depth Offset Between F322W2 and F444W}

There is some overlap between the NIRCam F322W2 and F444W bandpasses. In these observations, this overlap region extended from 3.89--3.95 \um\ and contained four depth measurements from the F322W2 observation, and four depth measurements from the F444W observation. Visual inspection of the spectra showed a noticeable offset between the F322W2 and F444W data within this overlap region in all three of our reductions. For our nominal \texttt{Pegasus} results the F322W2 data in the overlap region were $163\pm39$\,ppm higher than the F444W data. Both the \texttt{Eureka!} and \texttt{tshirt} spectra showed similar offsets.

We used the Spitzer \three and \four transit depths we measured as a part of the joint broadband fit to determine which of the two NIRCam spectra was likely at the correct overall transit depth. To do so, we estimated a ``Spitzer equivalent'' transit depth at \three\ using a portion of our F322W2 data and one at \four using our F444W data. We used the Spitzer/\textsc{irac} filter and instrumental throughput curves provided by IPAC to account for Spitzer's varying sensitivity across each bandpass appropriately.

\begin{figure*}[t!] 
\begin{center}                                 
\includegraphics[width=1.0\textwidth]{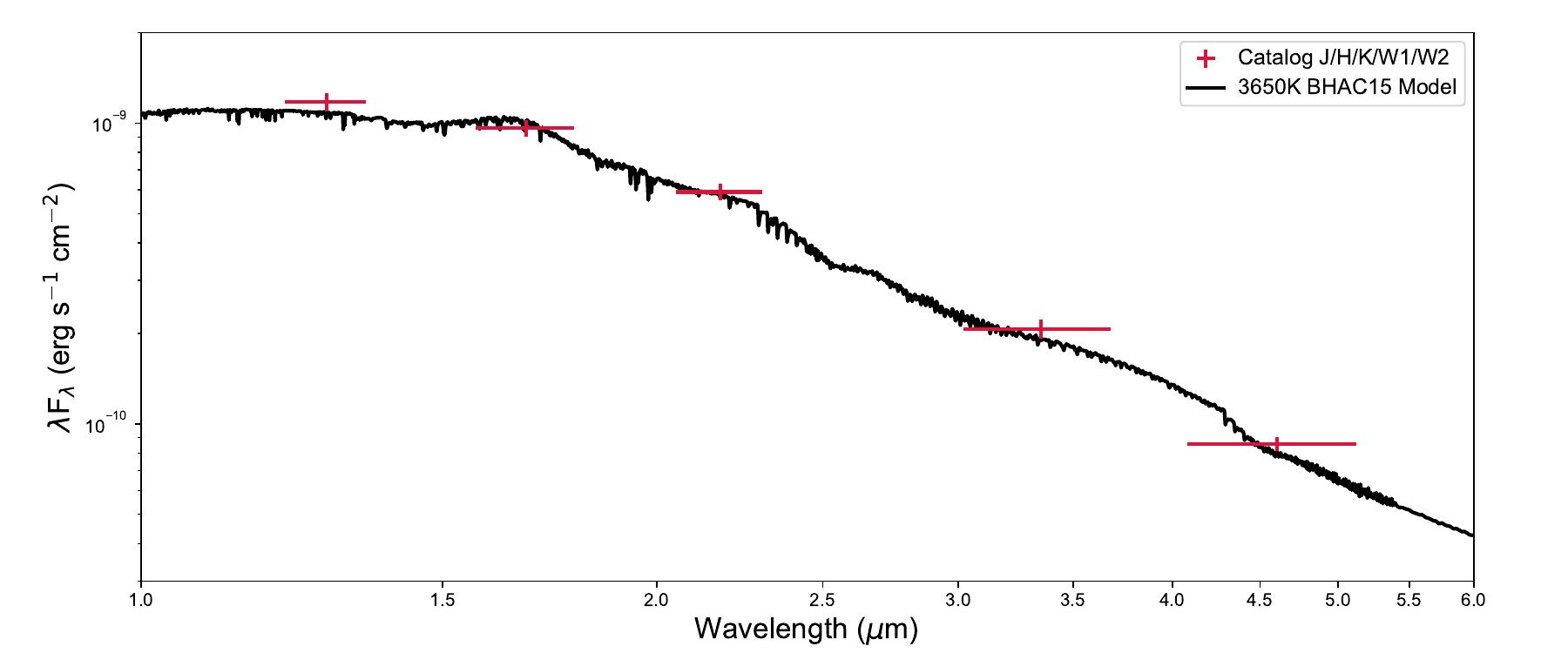}
\end{center}
\vskip -0.0in 
\caption{Our fit to the spectral energy distribution of GJ 3470. Using catalog 2MASS and AllWISE measurements of GJ 3470's brightness along with the Gaia DR3 parallax to the system we modeled the spectral energy distribution from the star GJ 3470. We measure the stellar radius to be $R_*=0.50\pm0.01$\,\rsun\ and its mass to be $M_*=0.44\pm0.04$\,\msun. Using the results from our lightcurve fitting, this allows us to refine the radius and mass of the planet GJ 3470 b to be $R_p=4.22\pm0.09$\,\re\ and $M_p=11.1\pm0.9$\,\me.}
\label{fig:sed}  
\end{figure*}

We find that the F444W spectrum almost perfectly matches the Spitzer \four transit depth, while the F322W2 spectrum is significantly higher than the Spitzer \three transit depth. Specifically, the F444W spectrum gives a ``\four equivalent'' depth that is lower by $17\pm39$\,ppm than the measured Spitzer \four depth. The F322W2 spectrum gives a ``\three equivalent'' depth that is higher by $144\pm33$\,ppm than the measured Spitzer \three depth. This comparison to an independent measurement indicates that the F444W spectrum is likely at the correct overall transit depth, and the F322W2 spectrum is offset high.

\subsection{HST Spectroscopic Transit Fitting}

To fit the spectroscopic transit depths in our re-reduction of the existing WFC3 G141 data, we used a \texttt{BATMAN} \citep{batman} transit model and standard detrending techniques for WFC3 timeseries data \citep{beatty2017kepler13}. As with our analysis of the JWST observations we fixed the transit center time, orbital period, scaled semi-major axis, orbital period, eccentricity, and the argument of periastron to the bestfit values from the joint broadband fit, and used a quadratic limb-darkening law with coefficients derived from the ATLAS9 \citep{atlas9} stellar models. We fit each spectroscopic channel individually, with the planet-to-star radius ratio and limb-darkening coefficients as the only free astrophysical parameters. We simultaneously fit the spectroscopic lightcurves from both visits for each channel, using the same value of $R_p/R_*$ and limb-darkening for both.

To model the spacecraft and detector systematic trends present in the WFC3 data, we used an exponential ramp within each individual orbit, and a linear temporal trend across each visit, of the form

\begin{equation}\label{eq:3310}
F_{detrend} = (m\,t_V+n)\,\left(1-A\,e^\frac{t_O}{\tau}\right).
\end{equation}

Here, $t_O$ is the time since the start of each orbit's observations, $t_V$ is the time since the start of the visit, and $A$ and $\tau$, and $m$ and $n$ are fitting coefficients for the exponential ramp and linear trend respectively. In addition to fitting the systematics in this way, we also did not use the data from the first orbit within each visit nor the first exposure from each orbit when fitting the data.

The measured transmission spectrum from our re-analysis of the WFC3 observations is given in Table 5.

\section{Stellar SED Fitting and Absolute Masses and Radii}

We used a set of catalog magnitudes for the GJ 3470 system to fit an SED model to the stellar emission. In conjunction with the Gaia \citep{GaiaDR3} parallax for the system, this allowed us to estimate a stellar radius for the star GJ 3470.

We used catalog 2MASS JHK \citep{2mass} and AllWISE W1 and W2 \citep{wise} magnitudes for our SED fits. Our SED model used four different physical parameters: the stellar effective temperature, the stellar radius, the amount of visual extinction to GJ 3470, and the system's parallax. We assumed $\log(g)=4.5$ and $\feh=0.0$ for the star GJ 3470, which are both within 0.2 dex of the surface gravity and metallicity measured by \cite{biddle2014} In practice, the precise values of $\log(g)$ and $\feh$ do not significantly affect the results from the SED fit \citep{stevens2018}.

We imposed a Gaussian prior on $\teff$ based on the spectroscopic measurements from\cite{biddle2014} of $\teff=3650\pm50$\,K with the associated $1\,\sigma$ uncertainties as the prior width. Similarly, we imposed a Gaussian prior on the parallax to the system using the Gaia DR3 \citep{GaiaDR3} parallax of $\pi=34.0172\pm0.0255$\,mas. For the amount of visual extinction to GJ 3470 we imposed a prior of $A_V=0.03\pm0.01$, which comes from measurements \citep{sf2011dustmap} of the excess reddening towards GJ 3470 of $E(B-V)=0.027\pm0.01$, and assuming $R_V=3.1$.

To model the SED, we used BHAC15 spectra \citep{bhac15} for the stellar SED. We computed a grid of surface luminosity magnitudes, corresponding to the bandpasses of the catalog magnitudes, for a range of $\teff$ values. Since the BHAC15 models step by 200K in $\teff$ we used cubic spline interpolation to estimate model magnitudes in between the points provided by the model atmospheres. We then scaled the interpolated surface magnitudes for the star by $R_*/d$ -- where $d$ is the distance to the star -- to determine the apparent bolometric flux of the SED at Earth. We then applied a simple $R=3.1$ extinction law scaled from the value of $A_V$, to determine the extincted bolometric flux of the SED model.

This stellar SED fitting (Figure \ref{fig:sed}) allowed us to measure the radius of the star GJ 3470 to be $R_*=0.50\pm0.01$\,\rsun. We then used the mean value of $R_P/R_*$ as measured in our joint broadband transit fits to estimate the planetary radius of GJ 3470 b to be $R_p=4.22\pm0.09$\,\re. Using the stellar density we measure as a part of the joint broadband transit fits, we constrain GJ 3470's stellar mass to be $M_*=0.44\pm0.04$\,\msun, and using a mass-ratio measured via radial velocity observations \citep{kosiarek2019}, we estimate a planetary mass of $M_p=11.1\pm0.9$\,\me. Our atmospheric retrievals use these absolute stellar and planetary radii and masses. 

\section{Stellar Variability}

\begin{figure*}[t!] 
\begin{center}                                 
\includegraphics[width=1.0\textwidth]{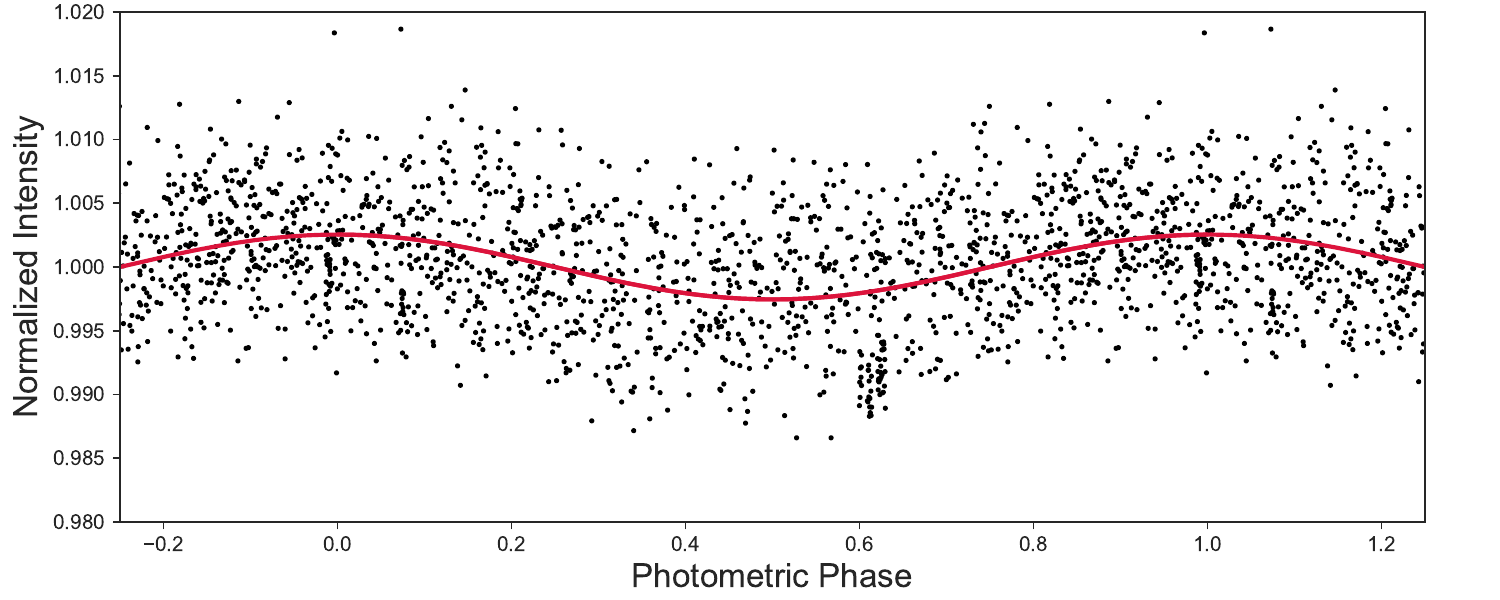}
\end{center}
\vskip -0.0in 
\caption{Photometric monitoring of the GJ 3470 system from 2012 to 2023 shows sinusoidal brightness modulations with an amplitude of $2545\pm130$\,ppm at a period of $P=21.6239\pm0.0014$\,days in a Cousins $R$ filter. The amplitude of stellar variability decreases significantly at longer wavelengths\cite{benneke2019gj3470} and we expect this variability to affect our measured transit depths at $<$10\,ppm over the F322W2 and F444W wavelength range.}
\label{fig:starvar}  
\end{figure*}

Previous photometric monitoring of GJ 3470 during the 2012-2013 season showed that the star itself is slightly variable at the presumed stellar rotation period of 21 days \citep{biddle2014}. This same monitoring campaign now spans from 2012 to 2023 and has given us a more precise measurement of stellar variability (Figure \ref{fig:starvar}). 

We observed the GJ 3470 system with the Tennessee State University Celestron 14-inch (C14) automated imaging telescope (AIT) located at Fairborn Observatory in southern Arizona \citep{ehf2003}. From 2012 to 2023 we collected 1376 good observations using a Cousins $R$ filter with an SBIG STL-1001E CCD camera. Each nightly observation consisted of three to ten consecutive exposures of the GJ 3470 field of view, which we use to perform differential aperture photometry on GJ 3470. Further details of the observing and data reduction procedures have been described previously in \cite{sing2015wasp31}.

Using these new additional observations, we find that the GJ 3470 system shows sinusoidal brightness modulations with an amplitude of $2545\pm130$\,ppm at a period of $P=21.6239\pm0.0014$\,days. This variability amplitude is roughly half the value measured previously using only the 2012-2013 data \citep{biddle2014} and roughly half the variability amplitude used in previous modeling of how stellar variability could affect GJ 3470 b's transmission spectrum in \cite{benneke2019gj3470}. Note that we have normalized the seasonal means of the photometric observations to all equal unity. This allows us to estimate the short-term (tens of days) variability, but not multi-year variability, of GJ 3470.

If we conservatively assume that the amplitude of GJ 3470's rotational variability is constant as a function of wavelength, then this will introduce at most a 30\,ppm offset between our various observations. In reality, the amplitude of stellar variability decreases significantly at longer wavelengths. \cite{benneke2019gj3470} modeled the wavelength dependence of GJ 3470's variability and its effect on transmission spectra using a variability amplitude approximately twice that which we measured, and found that it was at most 20\,ppm in the WFC3 bandpass and $<$10\,ppm over the F322W2 and F444W wavelength range. Since even our conservative estimate for the effect of stellar variability -- assuming a wavelength-independent amplitude -- is on the order of (or smaller) than our estimated depth uncertainties and the inter-instrument offsets, we do not expect the variability seen in the GJ 3470 system to affect our results significantly.  

\section{Atmospheric Modeling and Retrievals}
In order to derive quantitative atmospheric constraints we perform a series of atmospheric retrievals--Bayesian parameter estimation with a planetary atmosphere/spectral model.  Our setup/philosophy follows similar to that described in \cite{bell2023_methane} and Welbanks et al. submitted. Briefly, we perform both 1D-radiative-convective-photochemical equilibrium grid-based (1D-RCPE) retrievals and more flexible, classic, free-retrievals, each summarized below \citep{welbanks2021aurora}.

\subsection{1D-RCPE Grid-Retrieval}\label{sec:RCPERetrieval}
We use the {\tt ScCHIMERA} \citep{piskorz2018, mansfield2021, bell2023_methane} 1D-radiative-convective-photochemical-equilibrium solver (1D-RCPE) to produce a physically self-consistent grid of atmospheric models that are then fit to the spectra using the methods outlined in \cite{bell2023_methane}. The grid is a function of metallicity (0 $\le$ [M/H] $\le$ 3.0 in steps of 0.125), C/O (0.01$\le$C/O$\le$0.6, using 8 spacings), internal temperature (100$\le$T$_{\rm int}$$\le$350 in steps of 50 K), and a vertically constant eddy mixing (7$\le$log$_{10}K_{zz}$$\le$10 in steps of 0.5), at a fixed effective irradiation temperature (at the planetary equilibrium temperature--653 K). These resulted in 8400 converged 1D-RCPE model atmospheres. For the photochemistry/kinetics (using the VULCAN \citep{tsai2021vulcan} chemical kinetics code), we use the H-O-C-N-S reaction list \citep{Tsai2022} and the MUSCLES \citep{MusclesI,MusclesII,MusclesIII} GJ 176 spectrum as a proxy for GJ 3470. 

As in previous work \citep{bell2023_methane}, we fit for the metallicity, C/O, internal temperature, eddy diffusivity, planetary radius and planetary radius reference pressure, gray and power law patchy clouds \citep{welbanks2021aurora}, and stellar heterogeneity \citep{IyerLine2020}. Parameter estimation is performed with the {\tt pymultinest} \citep{Buchner2014} Nested Sampling tool. Within the Nested Sampling, the pre-computed  1D-RCPE model atmospheres (temperature, molecular weight, and gas volume mixing ratio profiles) are interpolated (using the python/scipy RegularGridInterpolator) to the sampled metallicity, C/O, internal temperature, eddy diffusivity. The interpolated model atmosphere at each step/sample--along with the current cloud, reference radius/pressure, and stellar heterogeneity parameters are used to compute a 1D high-resolution (R=100K) transmission spectrum.  This spectrum is then binned down to the data resolution to compute the chi-square log-likelihood used by the sampler. In summary, these grid-based constraints suggest a metallicity $125\pm40$ times Solar, a slightly sub-solar $\mathrm{C}/\mathrm{O}=0.35\pm0.1$, weak vertical mixing, and a poor constraint on the internal temperature.  A power law scattering slope is needed to explain the join the WFC3-to-NIRCam continuum/slope. Stellar heterogeneity also appears to play little role in the shape of the spectrum as the spot-covering fraction pushes towards the lower prior bound of zero. Main text Figure \ref{fig:abundances} shows the agreement between the molecular abundances retrieved from the free retrieval and those predicted by the retrieved grid-based retrieval parameters suggesting that the inferred compositional state is plausible. 

\subsection{Free-Retrievals}\label{sec:FREERetrieval}

We also explored GJ 3470 b's atmospheric properties using more flexible and agnostic forward models in a Bayesian inference procedure. These models, known as free-retrievals, are methods to retrieve the chemical abundances of different gases, the vertical temperature structure, and the cloud/haze properties on the planetary atmosphere. Compared to the 1D-RCPE Grid-Retrieval explained above, these methods do not assume any physico-chemical equilibrium conditions, and instead aim to capture the atmospheric conditions directly through a series of parameters for the chemistry and physical conditions of the atmosphere without expectations of physical consistency. These more flexible approaches provide the opportunity to capture conditions that otherwise would be prohibited by self-consistent models, such as combinations of gases not considered under chemical equilibrium. Nonetheless, caution must be exercised in the interpretation of free-retrievals as model assumptions may contribute to biased and unphysical atmospheric estimates \citep{welbanks2022terminators}. 

We employ the retrieval framework Aurora \citep{welbanks2021aurora} capable of modeling and interpreting transmission and emission spectra of transiting exoplanets \citep{welbanks2022terminators,bell2023_methane}. A detailed description of Aurora's transmission modeling approach is described in \cite{welbanks2021aurora}. Briefly, the atmospheric model solves radiative transfer for a parallel-plane atmosphere under hydrostatic equilibrium for a transmission geometry. The atmospheric model considers a one-dimensional model atmosphere spanning from $10^{-9}$ bar to 100 bar, divided into 100 layers uniformly spaced in logarithmic pressure space. The vertical temperature structure is parameterized following \cite{madhu2009retrievals}. The model simultaneously retrieves the reference pressure and reference radius for the planetary gravity of $\log (g_{p})=2.78$ cgs (Table \ref{table:starplanetprops}).

The atmospheric models assume uniform mixing ratios for H$_2$O, CH$_4$, NH$_3$, CO, CO$_2$, SO$_2$, and HCN using independent free parameters for each gas' volume mixing ratio. The presence of clouds and hazes is incorporated via a one-sector parameterization \citep{welbanks2021aurora}, which treats the combined spectroscopic effect of an optically thick cloud deck at a pressure P$_\mathrm{cloud}$ together with hazes following an enhancement to Rayleigh-scattering \citep{Lecavelier2008} in a linear combination with a cloud-free atmosphere \citep{line2016nonuniformclouds}. 

The Bayesian inference is performed using nested sampling \citep{Skilling2004,Feroz2009} via \texttt{PyMultiNest} \citep{Buchner2014} using 500 live points in the sampling {\bf with an evidence tolerance of 0.5 and a sampling efficiency of 0.8}. Each forward model in the sampling was calculated using line-by-line opacity sampling at a spectral resolution of 20,000 and then convolved and binned to the resolution of the observations. In total, the sampling is performed over 19 parameters: 7 molecular gases, 6 for the pressure-temperature structure of the planet, 1 for the reference pressure, 1 for the reference radius, and 4 for the presence of inhomogeneous clouds and hazes. 

Figure \ref{fig:contrib} shows the retrieved transmission spectrum and the contributions of the detected gases in GJ 3470 b's atmosphere. The retrieved volume mixing ratios from the free-retrieval are consistent with the gas profiles from the 1D-RCPE inferences, as shown in Figure \ref{fig:abundances}.

Both free retrievals confirm the detections of H$_2$O, CH$_4$, CO$_2$, and SO$_2$ with detection significances of $\sim4\sigma$ or greater. We note that the term ``detection significance'' refer to a model preference between a reference model considering all gases and nested models for which each individual species is removed in turn \citep{Benneke2013,welbanks2021aurora}. As such, any quoted detection significance is dependent on the choice of models and choice of model priors. For GJ 3470 b, the agreement in retrieved molecular abundances and confirmation of the strong evidence for the detection of these species regardless of model assumptions (e.g., free abundances vs. radiative convective-photochemical-equilibrium) confirms the robustness of the interpretation of the planet's spectrum. Table \ref{tab:free-retrieved} shows the retrieved parameters, priors, and detection significances where applicable. 

\begin{table*}
\begin{center}
\caption{Median Values and 68\% Confidence Intervals for the Joint Broadband Fit}\label{table:broadband}
\begin{tabular}{l l l}
\hline
Parameter & Description/Units & Value \\ \hline
System Parameters: & & \\
~~~$T_C$\dotfill &Transit time (\bjdtdb)\dotfill & $2459884.250655\pm0.000015$\\
~~~$\log(P)$\dotfill &Log orbital period (days)\dotfill & $0.523311956\pm1.8\times10^{-8}$\\
~~~$e\cos{\omega}$\textsuperscript{a}\dotfill & \dotfill & $0.01457\pm0.00075$\\
~~~$e\sin{\omega}$\dotfill & \dotfill & $0.00\pm0.03$\\       
~~~$\cos{i}$\dotfill & Cosine of inclination \dotfill & $0.022\pm0.016$\\ 
~~~$\log(a/R_{*})$\dotfill &Log semi-major axis ($R_{*}$) \dotfill & $1.15\pm0.01$\\
~~~$R_{WFC3}/R_{*}$\dotfill &WFC3 Radius ratio\textsuperscript{a} \dotfill & $0.07783\pm0.00023$\\
~~~$R_{F210M}/R_{*}$\dotfill &F210M Radius ratio \dotfill & $0.077033\pm0.000086$\\
~~~$R_{F322W2}/R_{*}$\dotfill &F322W2 Radius ratio\textsuperscript{$\dag$} \dotfill & $0.07746\pm0.00015$\\
~~~$R_{IRAC1}/R_{*}$\dotfill &IRAC1 Radius ratio \dotfill & $0.07670\pm0.00014$\\
~~~$R_{F444W}/R_{*}$\dotfill &F444W Radius ratio\textsuperscript{$\dag$} \dotfill & $0.07715\pm0.00018$\\
~~~$R_{IRAC2}/R_{*}$\dotfill &IRAC2 Radius ratio \dotfill & $0.07722\pm0.00015$\\
\hline
Broadband Transit Depths: & & \\
~~~$\delta_{WFC3}$\dotfill & WFC3 transit depth (ppm)\textsuperscript{b}\dotfill & $6057 \pm 32$ \\
~~~$\delta_{F210M}$\dotfill & F210M transit depth (ppm)\dotfill & $5943 \pm 30$ \\
~~~$\delta_{F322W2}$\dotfill & F322W2 transit depth (ppm)\textsuperscript{b}\dotfill & $5999 \pm 21$ \\
~~~$\delta_{IRAC1}$\dotfill & IRAC1 transit depth (ppm)\dotfill & $5883 \pm 22$ \\
~~~$\delta_{F444W}$\dotfill & F444W transit depth (ppm)\textsuperscript{b}\dotfill & $5951 \pm 28$ \\
~~~$\delta_{IRAC2}$\dotfill & IRAC2 transit depth (ppm)\dotfill & $5963 \pm 24$ \\
\hline
Derived Properties: & & \\
~~~$P$\dotfill &Orbital period (days)\dotfill & $3.33666001\pm1.4\times10^{-7}$\\  
~~~$i$\dotfill & Inclination (deg.)\dotfill & $88.7\pm0.1$\\
~~~$a/R_{*}$\dotfill &Semi-major axis in stellar radii\dotfill & $14.05\pm0.35$\\
~~~$b$\dotfill &Impact parameter\dotfill & $0.31\pm0.02$\\
~~~$T_{14}$\dotfill &Full transit duration (hours)\dotfill & $1.833\pm0.048$\\
~~~$e$\dotfill & Eccentricity\dotfill & $0.024\pm0.012$\\
~~~$\omega$\dotfill & Argument of Periastron (deg.)\dotfill & $347\pm62$ \\
\hline
\multicolumn{3}{l}{\textsuperscript{a}\footnotesize{The value of $e\cos{\omega}$ recovers the Gaussian prior we imposed during the fitting process, which is}}\\
\multicolumn{3}{l}{\footnotesize{itself based on previous secondary eclipse observations \citep{benneke2019gj3470}.}}\\
\multicolumn{3}{l}{\textsuperscript{b}\footnotesize{The broadband WFC3, F322W2, and F444W data were included in the joint broadband fitting}}\\
\multicolumn{3}{l}{\footnotesize{process, but we used the transmission spectra from these instruments in our atmospheric modeling.}}\\
\end{tabular}
\end{center}
\end{table*}

\begin{table}
\begin{center}
\caption{GJ 3470 b's NIRCam/F322W2 Transmission Spectrum}\label{table:f322w2}
\vspace{3mm}
\begin{tabular}{c c }
\hline
Wavelength (\um)  & Transit Depth (ppm)\\ \hline
$2.450 - 2.465$ & $5915\pm70$ \\
$2.465 - 2.480$ & $5882\pm65$ \\
$2.480 - 2.495$ & $5780\pm72$ \\
$2.495 - 2.510$ & $5877\pm68$ \\
$2.510 - 2.525$ & $5900\pm66$ \\
$2.525 - 2.540$ & $5939\pm67$ \\
$2.540 - 2.555$ & $5875\pm68$ \\
$2.555 - 2.570$ & $5900\pm63$ \\
$2.570 - 2.585$ & $5882\pm64$ \\
$2.585 - 2.600$ & $5981\pm60$ \\
... & .. \\
$3.800 - 3.815$ & $5838\pm55$ \\
$3.815 - 3.830$ & $5732\pm55$ \\
$3.830 - 3.845$ & $5834\pm56$ \\
$3.845 - 3.860$ & $5795\pm57$ \\
$3.860 - 3.875$ & $5871\pm53$ \\
$3.875 - 3.890$ & $5820\pm56$ \\
$3.890 - 3.905$ & $5833\pm54$ \\
$3.905 - 3.920$ & $5795\pm56$ \\
$3.920 - 3.935$ & $5824\pm57$ \\
$3.935 - 3.950$ & $5867\pm54$ \\
\hline
\multicolumn{2}{l}{\footnotesize{The full machine-readable version of this}}\\
\multicolumn{2}{l}{\footnotesize{table is available online.}}\\
\multicolumn{2}{l}{\footnotesize{Note: These values have been offset by}}\\
\multicolumn{2}{l}{\footnotesize{$-163$\,ppm as discussed in Section 2.4}}\\
\multicolumn{2}{l}{\footnotesize{and Appendix B.3.}}\\
\end{tabular}
\end{center}
\end{table}

\begin{table}
\begin{center}
\caption{GJ 3470 b's NIRCam/F444W Transmission Spectrum}\label{table:f444w}
\vspace{3mm}
\begin{tabular}{c c }
\hline
Wavelength (\um)  & Transit Depth (ppm)\\ \hline
$3.890 - 3.905$ & $5817\pm67$ \\
$3.905 - 3.920$ & $5851\pm58$ \\
$3.920 - 3.935$ & $5828\pm56$ \\
$3.935 - 3.950$ & $5815\pm52$ \\
$3.950 - 3.965$ & $5924\pm51$ \\
$3.965 - 3.980$ & $5916\pm54$ \\
$3.980 - 3.995$ & $6041\pm53$ \\
$3.995 - 4.010$ & $5921\pm51$ \\
$4.010 - 4.025$ & $5913\pm56$ \\
$4.025 - 4.040$ & $5887\pm49$ \\
... & ... \\
$4.820 - 4.835$ & $5897\pm92$ \\
$4.835 - 4.850$ & $5932\pm83$ \\
$4.850 - 4.865$ & $6084\pm84$ \\
$4.865 - 4.880$ & $5852\pm91$ \\
$4.880 - 4.895$ & $6026\pm94$ \\
$4.895 - 4.910$ & $5781\pm97$ \\
$4.910 - 4.925$ & $5997\pm92$ \\
$4.925 - 4.940$ & $5840\pm104$ \\
$4.940 - 4.955$ & $5943\pm106$ \\
$4.955 - 4.970$ & $5832\pm108$ \\
\hline
\multicolumn{2}{l}{\footnotesize{The full machine-readable version of this}}\\
\multicolumn{2}{l}{\footnotesize{table is available online.}}\\
\end{tabular}
\end{center}
\end{table}

\begin{table}
\begin{center}
\caption{GJ 3470 b's HST/WFC3 Transmission Spectrum}\label{table:wfc3}
\vspace{3mm}
\begin{tabular}{c c }
\hline
Wavelength (\um)  & Transit Depth (ppm)\\ \hline
$1.120 - 1.156$ & $5991\pm42$ \\ 
$1.156 - 1.192$ & $6032\pm42$ \\ 
$1.192 - 1.228$ & $5983\pm42$ \\ 
$1.228 - 1.264$ & $5953\pm43$ \\ 
$1.264 - 1.300$ & $5966\pm43$ \\ 
$1.300 - 1.336$ & $6006\pm42$ \\ 
$1.336 - 1.372$ & $6122\pm43$ \\ 
$1.372 - 1.408$ & $6097\pm42$ \\ 
$1.408 - 1.444$ & $6057\pm42$ \\ 
$1.444 - 1.480$ & $5992\pm41$ \\ 
$1.480 - 1.516$ & $5998\pm43$ \\ 
$1.516 - 1.552$ & $5960\pm42$ \\ 
$1.552 - 1.588$ & $5955\pm42$ \\ 
$1.588 - 1.624$ & $5931\pm42$ \\ 
$1.624 - 1.660$ & $5939\pm43$ \\ 
\hline
\end{tabular}
\end{center}
\end{table}

\begin{table*}
\begin{center}
\caption{Median Values and 68\% Confidence Intervals for the Stellar and Planetary Properties}\label{table:starplanetprops}
\begin{tabular}{l l l}
\hline
Parameter & Description/Units & Value \\ \hline
Stellar Properties: & & \\
~~~$R_{*}$\dotfill &Stellar radius (\rsun)\dotfill & $0.50\pm0.01$\\
~~~$\rho_{*}$\dotfill &Stellar density (cgs)\dotfill & $4.9\pm0.35$\\
~~~$M_{*}$\dotfill &Stellar mass (\msun)\dotfill & $0.44\pm0.04$\\
~~~$\log (g_{*})$\dotfill &Log surface gravity (cgs)\dotfill & $4.68\pm0.03$\\
\hline
Planetary Properties: & & \\
~~~$R_{p}$\dotfill &Planetary radius (\re)\dotfill & $4.22\pm0.09$\\
~~~$\rho_{p}$\dotfill &Planetary density (cgs)\dotfill & $0.82\pm0.06$\\
~~~$M_{p}$\dotfill &Planetary mass (\me)\dotfill & $11.2\pm0.9$\\
~~~$\log (g_{p})$\dotfill &Log surface gravity (cgs)\dotfill & $2.78\pm0.03$\\
\hline
\end{tabular}
\end{center}
\end{table*}

\clearpage

\bibliography{references}{}
\bibliographystyle{aasjournal}

\end{document}